\documentclass{aa}
\usepackage{txfonts}
\usepackage{graphicx}

\begin{document}

\title{First Microlensing Candidates From The MEGA Survey\thanks{Based
on observations made with the Isaac Newton Telescope operated on the
island of La Palma by the Isaac Newton Group in the Spanish
Observatorio del Roque de los Muchachos of the Instituto de
Astrofisica de Canarias} Of M31}

\author{Jelte~T.A.~de~Jong\inst{1}
	\and Konrad~H.~Kuijken\inst{2,1}
	\and Arlin~P.S.~Crotts\inst{3}
	\and Penny~D.~Sackett\inst{4}
	\and Will~J.~Sutherland\inst{5}
	\and Robert~R.~Uglesich\inst{3}
	\and Edward~A.~Baltz\inst{3}
	\and Patrick Cseresnjes\inst{3}
	\and Geza Gyuk\inst{6}
	\and Lawrence~M.~Widrow\inst{7}\\
({\bf The MEGA collaboration})}

\institute{Kapteyn Astronomical Institute, University of Groningen, PO Box 800, 9700 AV, Groningen, The Netherlands
	\and Sterrewacht Leiden, University of Leiden, PO Box 9513, 2300 RA, Leiden, The Netherlands
	\and Columbia Astrophysics Laboratory, 550 W 120th St.,
	Mail Code 5247, New York, NY 10027, United States
	\and Research School of Astronomy and Astrophysics, Australian
	National University, Mt. Stromlo Observatory, Cotter Road,
	Weston ACT 2611, Australia
	\and Institute of Astronomy, Madingley Rd, Cambridge CB3 0HA,  
	United Kingdom
	\and Department of Astronomy and Astrophysics, University of
	Chicago, 5640 South Ellis Avenue, Chicago, IL 60637, United States
	\and Department of Physics, Queen's University, Kingston, ON
	K7L 3N6, Canada
}

\offprints{Jelte~T.A.~de~Jong, \email{jdejong@astro.rug.nl}}

\date{Received 24 September 2003 / Accepted 24 September 2003}

\abstract{
We present the first \object{M31} candidate microlensing events from
the Microlensing Exploration of the Galaxy and Andromeda (MEGA)
survey. MEGA uses several telescopes to detect microlensing towards
the nearby Andromeda galaxy, M31, in order to establish whether
massive compact objects are a significant contribution to the mass
budget of the dark halo of M31. The results presented here are based
on observations with the Isaac Newton Telescope on La Palma, during
the 1999/00 and 2000/01 observing seasons. In this data set, 14
variable sources consistent with microlensing have been detected, 12
of which are new and 2 have been reported previously by the
POINT-AGAPE group. A preliminary analysis of the spatial and timescale
distributions of the candidate events support their microlensing
nature. We compare the spatial distributions of the candidate events
and of long-period variable stars, assuming the chances of finding a
long-period variable and a microlensing event are comparable. The
spatial distribution of our candidate microlensing events is more
far/near side asymmetric than expected from the detected long-period
variable distribution.  The current analysis is preliminary and the
asymmetry not highly significant, but the spatial distribution of
candidate microlenses is suggestive of the presence of a microlensing
halo.
}

\maketitle

\section{Introduction}
One of the astrophysical solutions to the galactic dark matter problem
would be the presence of a significant amount of undetected compact
objects in the halos of galaxies. These MACHOs (Massive Astrophysical
Compact Halo Objects) can be detected using gravitational
microlensing (Paczynski \cite{Pacz86}). 
According to gravitational lensing theory the
measured brightness of a background source will temporarily increase
if a massive compact object moves close enough through our line of
sight towards the background source.

During the last decade, the MACHO (Alcock et al. \cite{Alcock93}) and
EROS (Aubourg et al. \cite{Aubourg93}) collaborations have been
monitoring fields in the Large and Small Magellanic Clouds
(\object{LMC} and \object{SMC}) in order to detect such events. After
5.7 years of observing the MACHO group found 13-17 microlensing events
towards the LMC, and concluded that up to 20\% of the Milky Way dark
halo may consist of compact objects of mass 0.15 - 0.9 $M_\odot$
(Alcock et al. \cite{Alcock00}). The EROS collaboration has found 3
microlensing events towards the LMC and puts strong constraints on the
fraction of dark matter in the form of compact objects (Lasserre et
al. \cite{Lasserre00}). The results of both groups are, however,
consistent with $\sim$10\% of the dark halo mass consisting of compact
objects of $\sim$0.5$M_\odot$ (e.g. Milsztajn \cite{Milsztajn02}).

Looking for microlensing events in the nearby Andromeda galaxy (M31)
has several advantages over Magellanic Clouds searches (Crotts
\cite{Crotts92}; Baillon et al. \cite{Baillon93}). Because of the
geometry the microlensing optical depth can be up to ten times larger
in parts of M31.  In combination with the extremely high density of
background stars, this results in a highly enhanced microlensing
rate. Due to the high inclination of the disk of M31, it should be
possible to use microlensing to constrain the mass contribution of
compact objects to the dark halo (Baltz, Gyuk \& Crotts
\cite{Baltz03}; Gyuk \& Crotts \cite{GyukCrotts00}; Kerins et
al. \cite{Kerins01}). In the presence of a significant microlensing
halo, the microlensing rate should be asymmetric, with more
microlensing taking place towards the far side of the disk than to the
near side, because the line-of-sight through the halo is longer
towards the far side.

Besides these advantages, M31 microlensing also has some
problems. Because of the large distance the stars are faint and
generally unresolved from the ground. However, using special
techniques, it is possible to detect microlensing in M31 (e.g. Crotts
\& Tomaney \cite{CrottsTomaney96}; Paulin-Henriksson et
al. \cite{Paulin03}; Calchi Novati et al. \cite{Calchi03}).
Three collaborations are currently working on microlensing surveys in
M31, namely POINT-AGAPE (Auri\`ere et al. \cite{Auriere01}), WeCAPP
(Riffeser et al. \cite{Riffeser01}) and MEGA (Crotts et al. \cite{MEGA}).

The Microlensing Exploration of the Galaxy and Andromeda (MEGA)
collaboration has performed an intensive four-year survey of two large
fields in M31 plus extended baseline observations in order to measure
the microlensing optical depth due to a possible MACHO halo. In this
paper we present the first microlensing candidates resulting from the
analysis of the 1999-2000 and 2000-2001 season data obtained at the
Isaac Newton Telescope at La Palma. In section 2 we briefly describe
the dataset and methods used. The microlensing candidates are presented in
section 3. Our discussion and conclusions are presented in section
4. Finally, future work is outlined in section 5.

\section{Data and Method}

\subsection{Dataset}

\begin{figure}
\centering
\includegraphics[width=8cm,clip=]{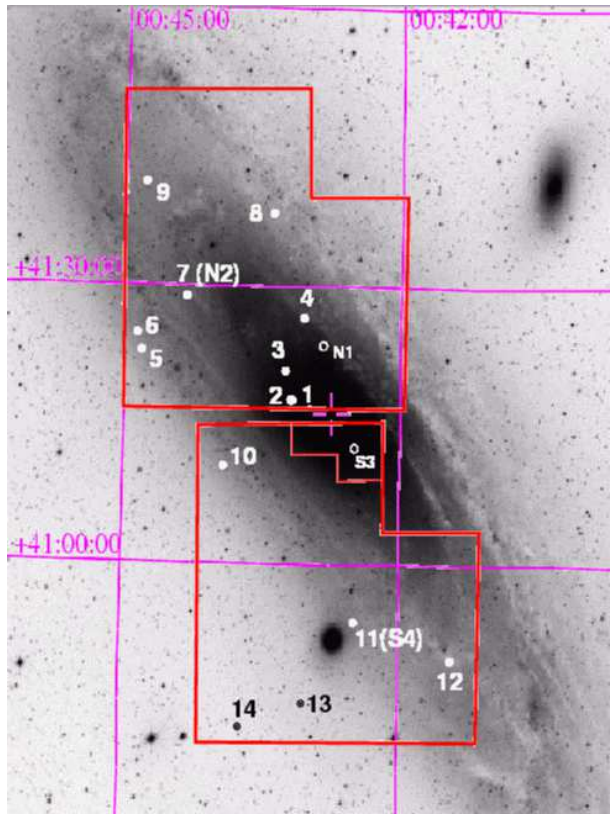}
\caption{The layout of the two INT Wide Field Camera (WFC) fields in
M31 used for the microlensing search described in this paper. The WFC
has four 2048x4100 pixel chips with a pixel scale of 0.333\arcsec,
offering a field of view of approximately 0.25$\Box^o$. With this
field layout we cover a large part of the far side (SE) of the disk of
M31 as well as part of the near side. Also plotted are the positions
of the candidate microlensing events in our sample (full
circles). Note that our events 7 and 11 correspond to the events
PA-99-N2 and PA-00-S4 reported previously by POINT-AGAPE. Part of the
south field close to the bulge is not used for our analysis, since the
image subtraction is not of high quality in this very high surface
brightness area. This region is indicated in the uppermost chip of the
south field. The two POINT-AGAPE events not present in our sample are
indicated with the open circles.}
\label{fig:candpos}
\end{figure}

MEGA uses several telescopes to monitor two wide fields, covering a
total area of 0.57 square degrees. For the current analysis,
observations done with the Wide Field Camera (WFC) on the Isaac Newton
Telescope (INT) were used. The layout of the WFC chips on M31 is shown
in Figure \ref{fig:candpos}. We present microlensing events
from the 1999/2000 and 2000/2001 observing seasons. The observations
in the first season (99/00) were spread over 57 nights between August
1st 1999 and January 4th 2000 and were taken in the Sloan
r$^{\prime}$, g$^{\prime}$ and i$^{\prime}$ broad bands. The
r$^{\prime}$ dataset is the largest with 57 epochs, whilst the
g$^{\prime}$ and i$^{\prime}$ datasets contain 41 and 24 epochs
respectively. During the second observing season (00/01) 90 epochs
were obtained in both r$^{\prime}$ and i$^{\prime}$, between August
1st 2000 and January 23 2001. Observations were spread equally over
both fields. The exposure time per epoch ranges between 5 and 30
minutes and is typically 10 minutes. Because the WFC is not always
mounted on the INT, the epochs tend to cluster in blocks of two to
three weeks. Part of the data from the 2001/2002 season has been used as
well. From this third season, 19 epochs in r$^{\prime}$ and
i$^{\prime}$ taken between August 13th 2001 and November 18th 2001
were used to extend the baselines of the events from the first two
seasons in order to exclude long period variable stars. These data
were not used to detect more microlensing events.

\subsection{Data reduction}

Standard data reduction, including bias subtraction, trimming and
flatfielding was performed in IRAF.
Because of the high stellar density
in M31 and its large distance, the background source stars are
usually resolved only while they are being lensed and sufficiently
magnified. To detect microlensing events in these fields, we use the
Difference Image Photometry (DIP) method as described by Tomaney \& Crotts
(\cite{TomaneyCrotts96}). This method involves subtracting individual
images from a high quality reference image, resulting in difference
images in which variable objects show up as residuals.

Below we outline the steps of the DIP pipeline that was used. All 
operations are done in IRAF, using
the DIFIMPHOT package written primarily by Austin Tomaney, unless
explicitly stated otherwise.\\
$\bullet$ {\it Astrometric registration and stacking of images}\\
All images are transformed to a common astrometric reference
frame. By stacking high quality images from the 1999 season, a high
signal-to-noise reference image was made. Per night all exposures are
combined separately for each band. Each epoch corresponds to the
combination of all frames taken in the same band in one night. The
Julian date of the epoch is taken as the weighted average of the
Julian dates of the individual frames.\\
$\bullet$ {\it Image subtraction}\\
From the single epoch
images the high signal-to-noise reference image is subtracted, after
photometric calibration and matching of the point spread function (PSF)
between the images (Tomaney \& Crotts \cite{TomaneyCrotts96}). 
The shape of the PSF is measured from bright, unsaturated stars in the
images that are being matched. By dividing the PSFs in Fourier space a
convolution kernel is calculated with which the better seeing image
(usually the reference image) is degraded.\\
$\bullet$ {\it Variable object detection}\\
The resulting difference
images are dominated by shot noise in which variable sources show up 
as positive or negative residuals, depending on the flux difference of
the object between the single epoch image and the reference
image. Due to fringing, the i$^{\prime}$ difference images are of
poorer quality than the r$^{\prime}$ difference images.
SExtractor (Bertin \& Arnouts \cite{SExtractor}) is used to detect residuals in all r$^{\prime}$
difference images from the first two observing seasons. The catalogs 
with residuals are cross-correlated
to obtain a catalog with all variable objects in the surveyed
fields. As a first selection to get rid of noisy detections, we demand
that objects have to be detected in at least two epochs.\\
$\bullet$ {\it Lightcurves and Epoch quality}\\
Lightcurves for the variable sources are obtained by performing PSF
fitting photometry on the residuals in the difference images, using
the PSF shape measured from the bright unsaturated stars.
Several epochs turned out to give problematic difference images for a
number of different reasons. Epochs with seeing worse than 2.0\arcsec
~do not give clean difference images and were discarded. A small number
of epochs were overexposed and had to be discarded as well.\\
Lightcurves were also produced at ``empty'' positions, i.e. positions
were no variability was detected. Flat line fits were done to these
empty lightcurves to check the error bars on the fluxes derived from
statistics of the PSF fitting photometry. For each epoch, the
distribution of the deviations from the flat lightcurve fits weighted
by the error bar returned by the photometry routine was examined. In
some cases this distribution showed broad non-gaussian wings, and
these epochs were discarded. Typically they were associated with
highly variable seeing between the individual exposures.  In other
cases, the normalized error distribution was gaussian, but with
dispersion higher than one. In these cases the error bars were
renormalized appropriately.

The typical number of epochs that were left after the procedure
described above are tabulated in table \ref{tab:epochs} for each filter
and field for the 99/00 and 00/01 seasons. 
From these epochs the lightcurves were constructed
that were used for the analysis presented in this paper.

\begin{table}
\begin{center}
\begin{tabular}{l|cccc}
\hline
\hline
~ & r$^{\prime}$ & ~ & i$^{\prime}$ & ~\\
~ & F1 & F2 & F1 & F2\\
\hline
99/00 & 46 & 45 & 16 & 15\\
00/01 & 56 & 55 & 56 & 56\\
\hline
Total & 102 & 100 & 72 & 71\\
\hline
\end{tabular}
\begin{small}
\caption{Overview of the number of epochs used for seasons 1 and 2, 
field and filter. F1 is the north field and F2 is the south field.}
\label{tab:epochs}
\end{small}
\end{center}
\end{table}

\subsection{Event selection}

The final dataset consists of lightcurves of 118,424 variable sources,
practically all of which are periodic variable stars. Finding
candidate microlensing events in such a large number of lightcurves is
no trivial problem. A procedure to select lightcurves that are
compatible with microlensing must be aimed at recognizing the
characteristics of a microlensing lightcurve while taking into account
computing speed, and the quality of the available data.  Since the
quality of the i$^{\prime}$ band data was clearly poorer than the
quality of the other bands, and since g$^{\prime}$ data were only
available for the first season, we decided to use the r$^{\prime}$
data as the main basis for candidate selection. Another advantage of
the r$^{\prime}$ data is that they have much better sampling density
in the first season than do i$^{\prime}$ and g$^{\prime}$ data
separately. Thus, the first steps of the filtering process involved
only r$^{\prime}$ data, after which the i$^{\prime}$ data are used to
further analyse the r$^{\prime}$ microlensing candidates.

A microlensing event caused by a single lens has a characteristic
shape and a flat baseline. The selection steps based purely on the
r$^{\prime}$ lightcurves are aimed at selecting lightcurves with flat
baselines and well sampled, significant peaks which fit the
characteristic microlensing shape well. A detailed description of
these steps is given in Appendix A1. To make sure these selection steps
do a good job selecting microlensing events and rejecting long-period
variable stars, Monte Carlo simulations of microlensing events and
long-period variables were performed and used to fine tune the
selection procedure. Based on just the r$^{\prime}$ data we
select 1,347 lightcurves.

Colour information can be used for the microlensing candidate
selection for several reasons. Contrary to variable stars,
microlensing events are intrinsically achromatic, meaning that the
colour of the observed difference flux remains the same during the
event. Furthermore, long-period variable stars that might be mistaken
for microlensing are very red. Because of this, the brightness
variations are also more pronounced in i$^{\prime}$ than in
r$^{\prime}$. Further selection steps, described in more detail in
Appendix A2, include a colour cut designed to
reject long-period variables and a goodness of fit criterion for the
i$^{\prime}$ lightcurve. For the latter, the i$^{\prime}$ lightcurves are
fit with a standard microlensing lightcurve with the shape parameters
found for the r$^{\prime}$ lightcurves, as a test for achromaticity.

\begin{figure*}
\centering
\includegraphics[width=15cm]{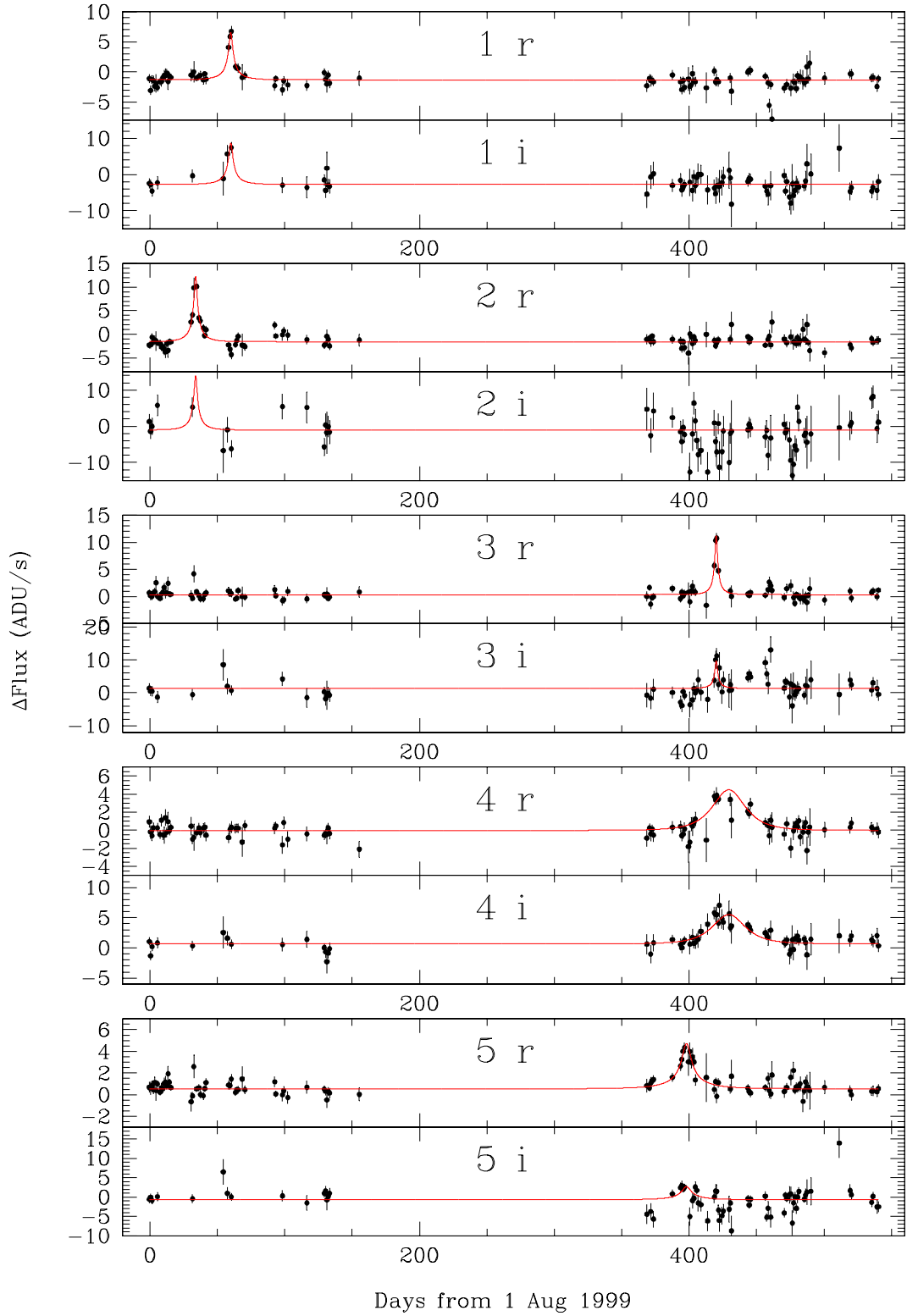}
\caption{Lightcurves of the 14 microlensing candidate events. This
page shows candidates 1 through 5, with the r$^{\prime}$
lightcurves plotted in the upper panel and the i$^{\prime}$
lightcurves in the lower panel. The Julian Date of the zeropoint of
the time axis is 2451393, which corresponds to August 1st 1999.}
\label{fig:candidates}
\end{figure*}

\setcounter{figure}{1}
\begin{figure*}
\centering
\includegraphics[width=15cm]{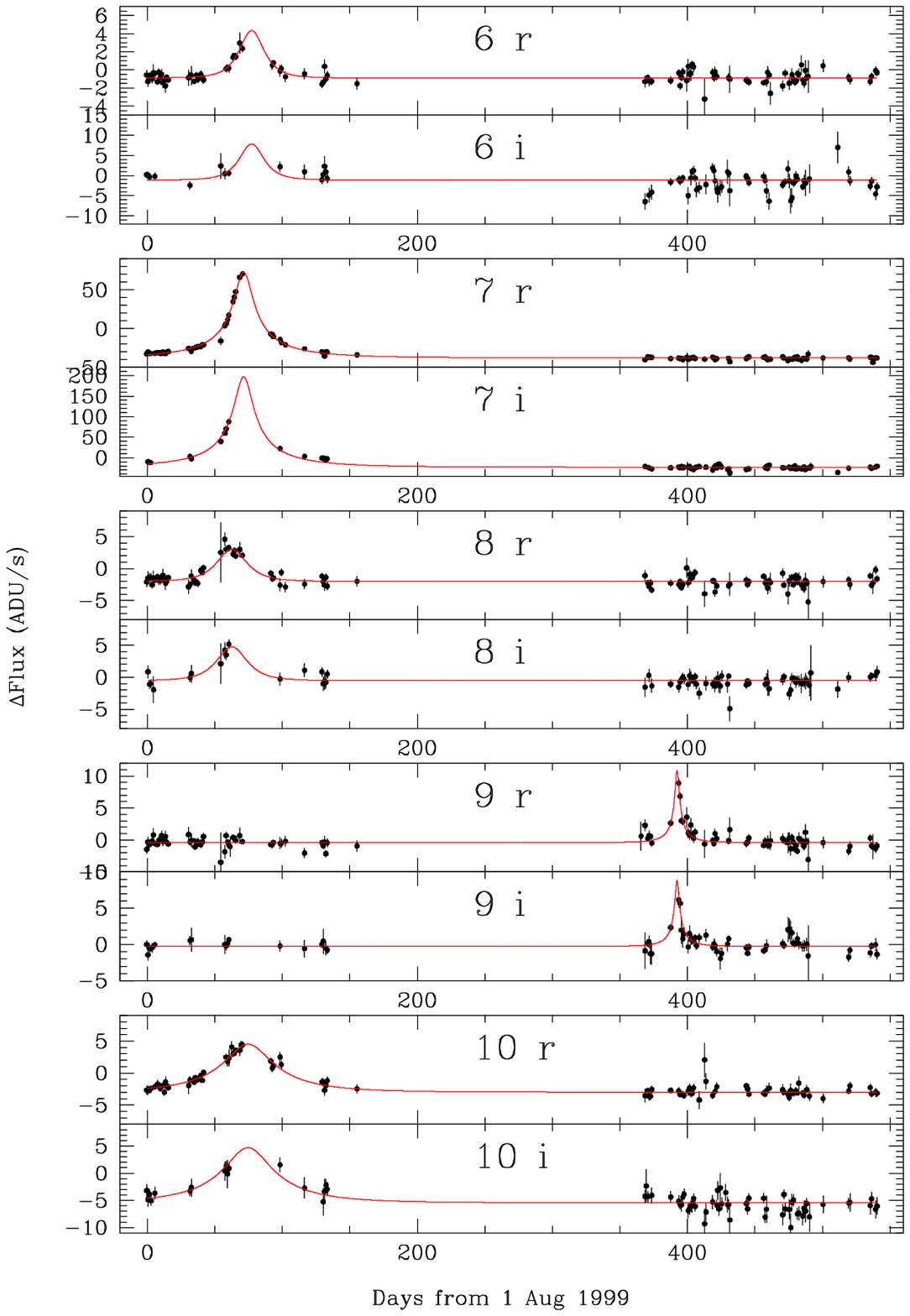}
\caption{continued -- Candidates 6 through 10.}
\end{figure*}

\setcounter{figure}{1}
\begin{figure*}
\centering
\includegraphics[width=15cm]{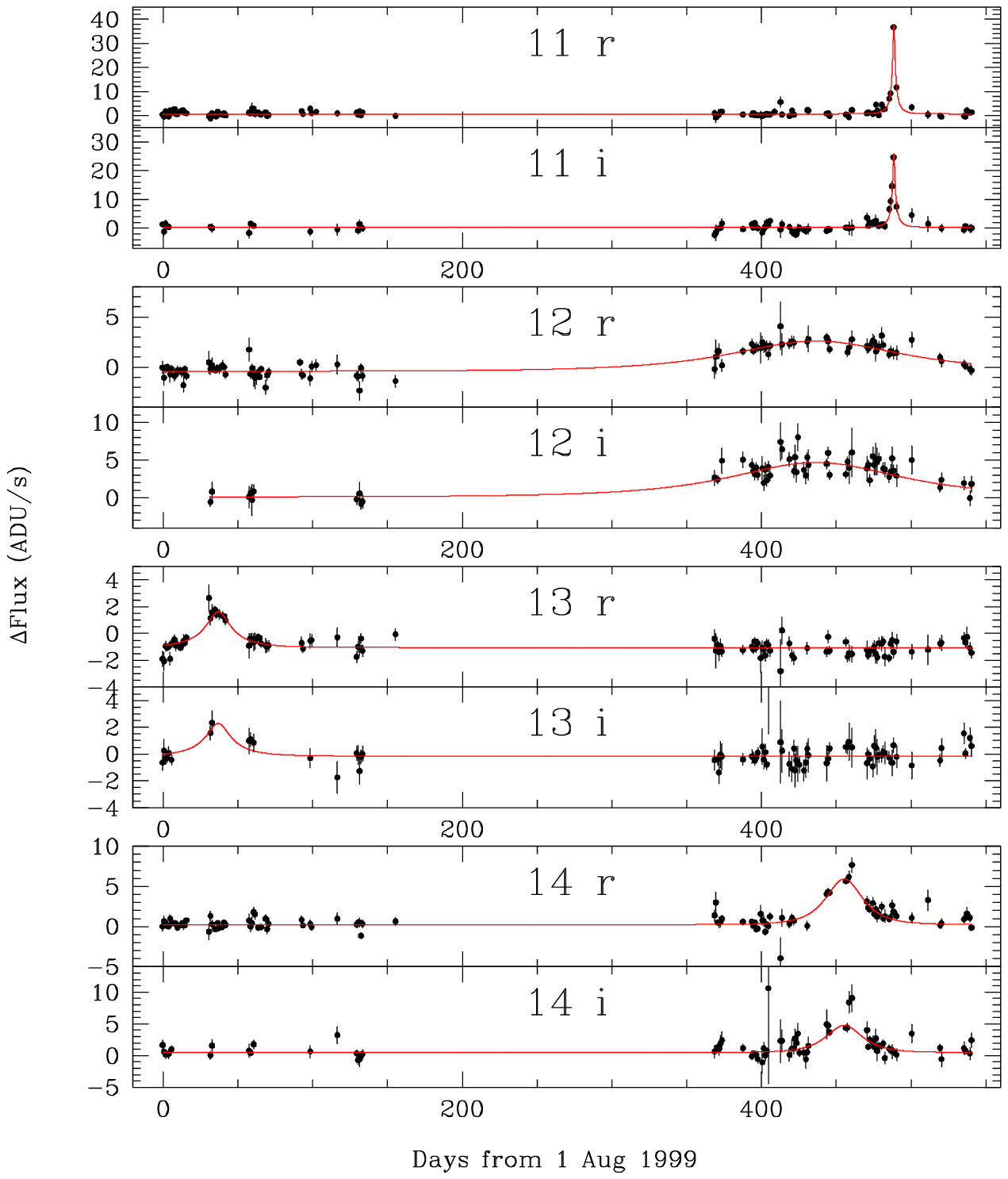}
\caption{continued -- Candidates 11 through 14.}
\end{figure*}

This complete automated selection procedure results in 126 candidate
microlensing events from the 99/00 and 00/01 observing seasons.
Unfortunately, rather lenient $\chi^2$ cuts had to be used for the
goodness of fit criteria. This is due to the
photometry being too sensitive to nearby variable sources, as
described in more detail in Appendix A. Because of this, the sample
still contains lightcurves with slightly variable baselines. The
lightcurves and the difference image residuals 
of these candidate events had to be inspected visually to establish
whether any additional variability was intrinsic to the variable source
or was caused by nearby sources. An effort to improve the photometry in the
presence of neighbours is underway.
The lightcurves with baseline variability that could not be attributed to
nearby variable sources, in total 67, were thrown out. In a further 39 cases,
the detected candidate event was actually caused by bad pixels and
these were discarded as well, so after visual inspection, 21 events remain that are
consistent with microlensing based on the 2 year lightcurves. Finally,
part of the data from the third (01/02) season were used to extend the
baselines of these 21 events, 7 of which showed variability in the
third season. The final set of candidate microlensing events consists
therefore of the 14 events described in the following section.

\section{Results}

\begin{table*}
\begin{center}
\begin{tabular}{l|llrrcc}
\hline
\hline
Candidate & RA & DEC & $t_{max}$ & $t_{FWHM}$ & $\Delta$r$^{\prime}$ & r$^{\prime}$-i$^{\prime}$\\
event & (J2000) & (J2000) & (days) & (days) & (mag) & (mag) \\
\hline
MEGA-ML 1 & 0:43:10.54 & 41:17:47.8 & 60.1 $\pm$ 0.1 & 4.2 $\pm$ 4.3 &
22.2 $\pm$ 1.1 & 1.1 $\pm$ 1.5\\
MEGA-ML 2 & 0:43:11.95 & 41:17:43.6 & 34.08 $\pm$ 0.08 & 4.6 $\pm$ 0.6 &
21.6 $\pm$ 0.3 & ~ \\
MEGA-ML 3 & 0:43:15.76 & 41:20:52.2 & 420.06 $\pm$ 0.05 & 2.6 $\pm$ 2.2 &
21.8 $\pm$ 1.2 & 0.4 $\pm$ 1.5\\
MEGA-ML 4 & 0:43:04.08 & 41:26:15.6 & 429.7 $\pm$ 0.2 & 29.1 $\pm$ 1.0 &
22.8 $\pm$ 0.2 & 0.8 $\pm$ 0.3\\
MEGA-ML 5 & 0:44:48.95 & 41:22:59.3 & 398.1 $\pm$ 0.2 & 9.4 $\pm$ 4.1 &
22.9 $\pm$ 0.8 & 0.4 $\pm$ 1.0\\
MEGA-ML 6 & 0:44:50.97 & 41:24:42.4 & 77.3 $\pm$ 0.2 & 22.9 $\pm$ 0.7 &
22.6 $\pm$ 0.2 & ~\\
MEGA-ML 7 & 0:44:20.89 & 41:28:44.6 & 71.20 $\pm$ 0.06 & 21.6 $\pm$ 0.7 &
19.3 $\pm$ 0.2 & 1.4 $\pm$ 0.2\\
MEGA-ML 8 & 0:43:24.53 & 41:37:50.4 & 63.1 $\pm$ 0.2 & 27.4 $\pm$ 0.9 &
22.7 $\pm$ 0.2 & 0.8 $\pm$ 0.2\\
MEGA-ML 9 & 0:44:46.80 & 41:41:06.7 & 392.3 $\pm$ 0.2 & 3.8 $\pm$ 1.6 &
21.8 $\pm$ 0.8 & 0.4 $\pm$ 1.0\\
MEGA-ML 10 & 0:43:54.87 & 41:10:33.3 & 74.7 $\pm$ 0.4 & 46.8 $\pm$ 4.4 &
22.2 $\pm$ 0.3 & 1.0 $\pm$ 0.3\\
MEGA-ML 11 & 0:42:29.90 & 40:53:45.6 & 488.6 $\pm$ 0.05 & 2.0 $\pm$ 0.3 &
20.5 $\pm$ 0.2 & 0.3 $\pm$ 0.2\\
MEGA-ML 12 & 0:41:26.90 & 40:49:42.1 & 436.6 $\pm$ 0.6 & 131.0 $\pm$ 9.4
& 23.2 $\pm$ 0.3 & 1.0 $\pm$ 0.3\\
MEGA-ML 13 & 0:43:02.49 & 40:45:09.2 & 37.3 $\pm$ 0.5 & 22.8 $\pm$ 3.8 &
23.3 $\pm$ 0.3 & 0.5 $\pm$ 0.4\\
MEGA-ML 14 & 0:43:42.53 & 40:42:33.9 & 455.1 $\pm$ 0.3 & 28.1 $\pm$ 1.4 &
22.5 $\pm$ 0.2 & 0.3 $\pm$ 0.4\\
\hline
\end{tabular}
\begin{small}
\caption{Fit parameters for the 14 candidate microlensing events. 
The $t_{max}$ and $t_{FWHM}$ values are based on full
parameter fits to the r$^{\prime}$ lightcurves. Also listed are the
magnitude of the difference flux at maximum amplification and the
r$^{\prime}$-i$^{\prime}$ colours where possible. For the short
events, i.e. $t_{FWHM}$ in the order of a few days, the uncertainties
are large, because the time sampling of the data is relatively sparse
compared to the width of the peak.
Peak times are in days after August 1st 1999.}
\label{tab:fitpars}
\end{small}
\end{center}
\end{table*}

\begin{figure}
\centering
\includegraphics[width=8cm]{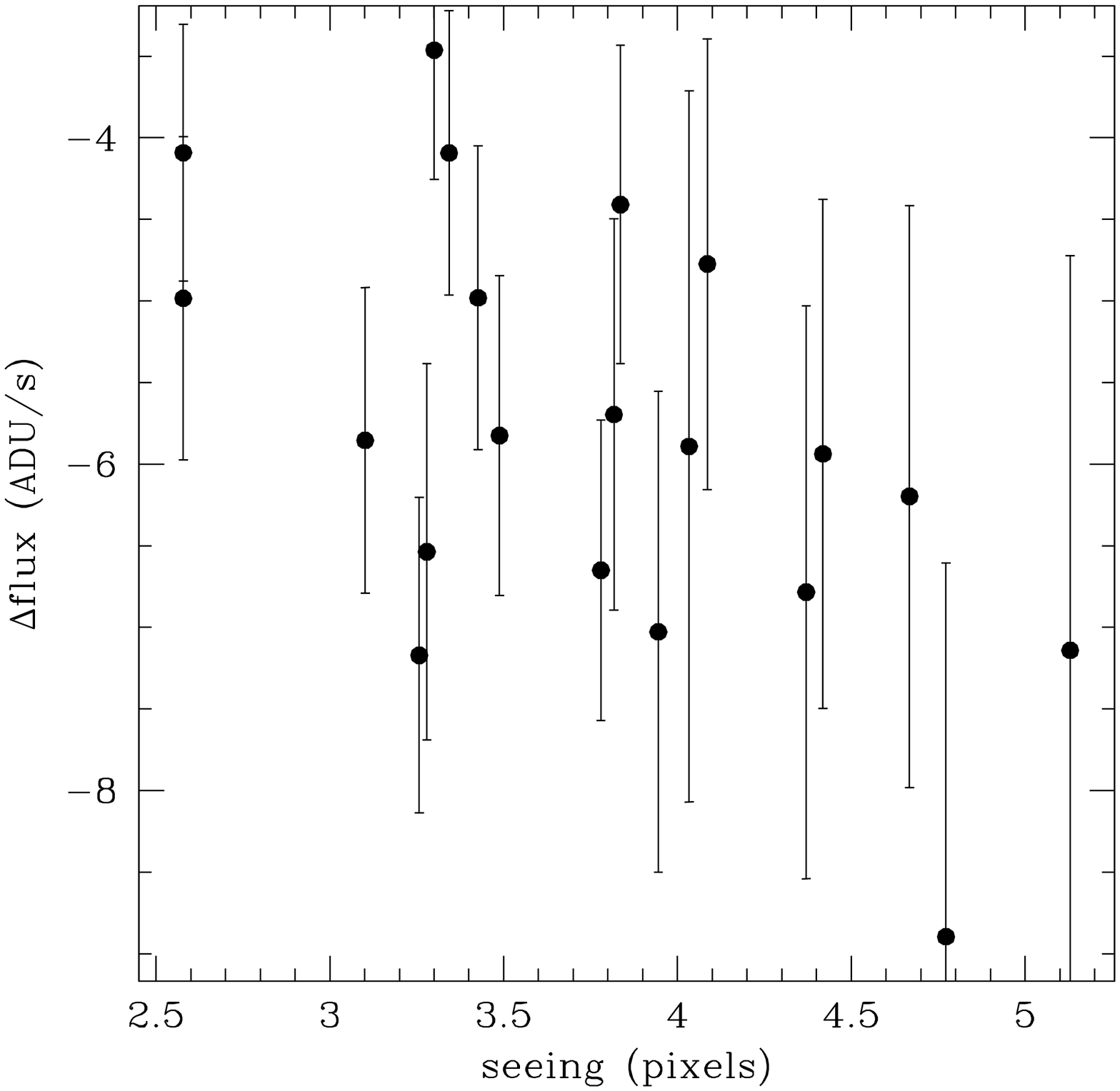}
\caption{Data points from the baseline of event 10, but now the
$\Delta Flux$ is plotted as function of the seeing. During the period
from which these points are taken, a variable source located about 7
pixels away from the event caused a strong residual in the difference
images. There clearly is a trend with worse seeing corresponding to
lower fluxes.}
\label{fig:seeingflux}
\end{figure}

\begin{figure*}
\centering
\includegraphics[width=15cm,angle=-90]{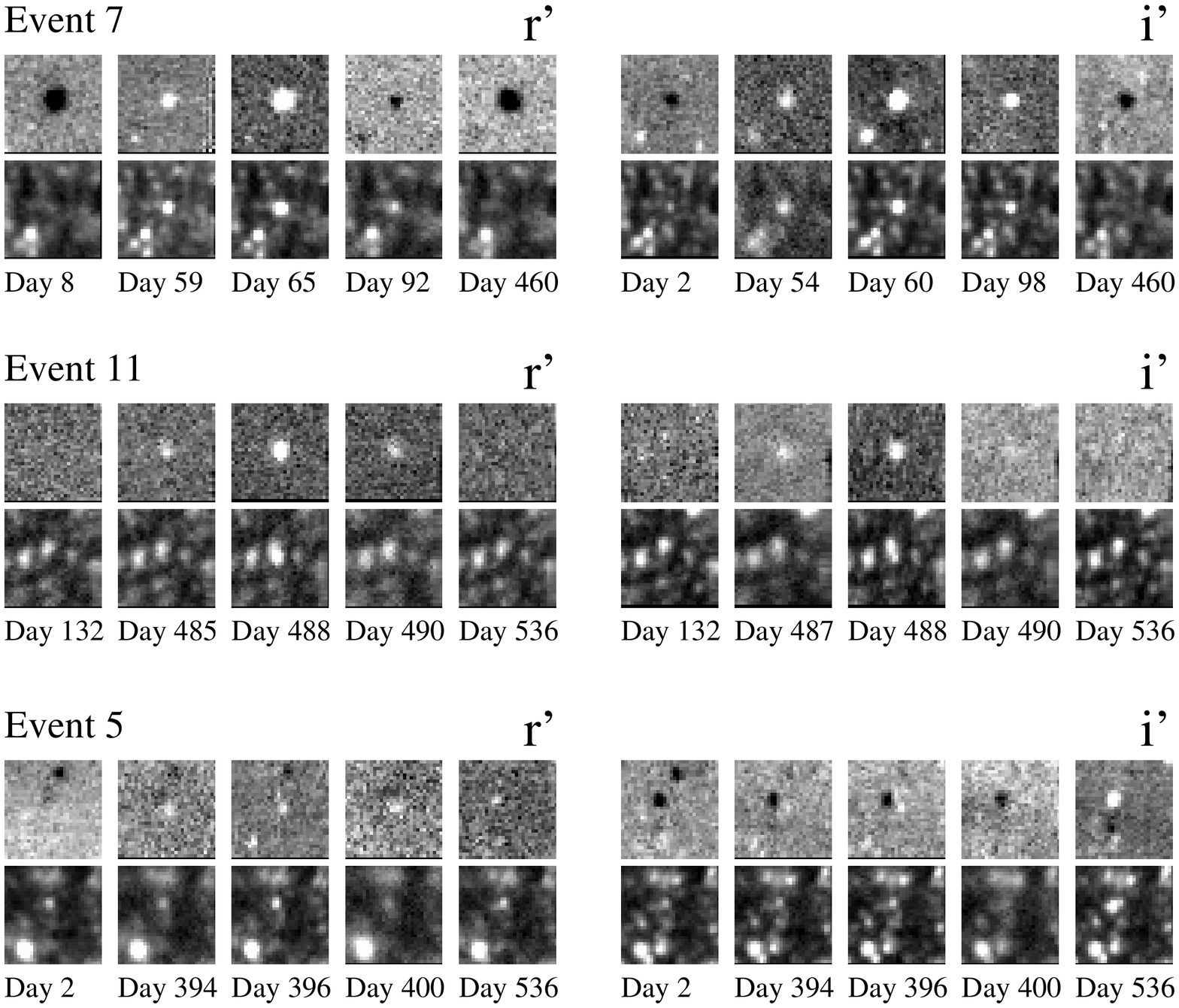}
\caption{Thumbnails of three events, chosen to span a wide range of
peak fluxes. The uppermost panel shows event 7, by far the brightest
event, going down in flux through event 11 (middle panel) to the low
signal-to-noise event 5 (lower panel). 
For each event we show thumbnails of the difference
images and single epoch images in both r$^{\prime}$ and in
i$^{\prime}$. Thumbnails were chosen from three epochs during the
microlensing event, i.e. in the peak of the lightcurve, and from one
epoch before and one epoch after the event.
Event 7 (upper panel) stands out very clearly in both the difference
images and the single epoch images in both bands. In the i$^{\prime}$
thumbnails a variable star is clearly visible in the lower left corner.
The event in the middle panel, event 11, has a peak flux only one
third of the peak flux of event 7, but still stands out clearly in the
difference images. In the single epoch images, the event is only
visible during maximum amplification.
Event 5, shown in the lower panel is an example of the fainter
events, with a peak flux almost an order of magnitude smaller than event
11. As can also be seen from the lightcurves (figure
\ref{fig:candidates}), the event is much brighter in r$^{\prime}$ than
in i$^{\prime}$. In the r$^{\prime}$ difference images the residual is
more obvious than in the i$^{\prime}$ difference images. The nearby
variable source that influences the i$^{\prime}$ photometry is clearly
visible just above and to the left of the event. In the single epoch
images, the event is hardly visible at all without difference image
techniques.}
\label{fig:thumbs}
\end{figure*}

The lightcurves of the 14 candidate microlensing events are shown in
figure \ref{fig:candidates}; their positions are
plotted in figure \ref{fig:candpos}. August 1st 1999 was taken as the
zero point of the time scale in the lightcurve plots.
The important fit parameters are tabulated in table \ref{tab:fitpars}.
For illustration we show in figure \ref{fig:thumbs} thumbnails of the
difference images and single epoch images for events 5, 7 and 11. All
difference image thumbnails for all events are available on-line at
http://www.astro.rug.nl/$\sim$jdejong/eventtn.

Inspection of the lightcurves in figure \ref{fig:candidates} indicates
that the
baselines are not always completely flat, this is especially the case
for the i$^{\prime}$ lightcurves. When looking at the difference images, it
becomes clear that for these events, secondary bumps in the lightcurve
are caused by variable objects very close to the position of the
event, rather than by variability in the events themselves. For
example, the bump around day 100 in the lightcurve of candidate event
number 2, is caused by a brightening source located only 4 pixels away
from the event. In the i$^{\prime}$ data, this problem occurs more frequently
because of the higher density of detected variable sources and the
effect is stronger because of the stronger variability at longer
wavelengths.

Not only can nearby variable sources result in trends in
the lightcurves, they can also increase the scatter in the flux. The
size of the aperture and the ring that is used for sky subtraction
depends on the seeing, meaning that a nearby variable does not
always influence the photometry and also not necessarily in the same
way. As an example of this, we show in figure \ref{fig:seeingflux} the
flux as a function of the seeing for the part of the baseline of event
10. Clearly, differences in seeing between the epochs will create
extra scatter in the baseline.

Below we discuss the possible reasons for additional
variability in each of the event lightcurves.\\ 
\noindent{\it Event 1}: This event has a bright periodic variable star at a distance
of 5 pixels, which is probably the cause of the noisy baseline in both
r$^{\prime}$ and i$^{\prime}$.

\noindent{\it Event 2}: This event also has close variable neighbours. In the r$^{\prime}$
difference images a variable source is located at only 3 pixels from
the event. This source peaks around day 100, causing a bump in the r$^{\prime}$
lightcurve. In the i$^{\prime}$ difference images there are two additional bright variables, both of them at about 10 pixels
distance. These sources cause the noisy nature of the i$^{\prime}$ baseline.

\noindent{\it Event 3}: Although the r$^{\prime}$ baseline is very clean, the i$^{\prime}$ baseline shows
quite some variability, especially in the second season. Again this is
caused by a very close variable source, at 3 pixels
distance. This source has a bright episode between days 440 and 470,
which causes the high fluxes in this period in the i$^{\prime}$ lightcurve.

\noindent{\it Event 4}: The baselines of this event look quite flat,
but the peak in the i$^{\prime}$ data is not very symmetric. The
sudden drop in flux after day 420 coincides with the sharp brightening
of a bright variable source at 14 pixels distance.

\noindent{\it Event 5}: This event has several variable sources
nearby. Unfortunately the closest one, at about 4 pixels, is also the
brightest, which especially in the i$^{\prime}$ lightcurve causes a
lot of noise. See figure \ref{fig:thumbs} for thumbnails of this event.

\noindent{\it Event 6}: In the second season the baseline shows a periodic
variability, which is caused by a periodic variable star at 6 pixels
from the event. A second close variable, at 5 pixels, shows up only in
the first season in r$^{\prime}$, but pops up also in the second season in i$^{\prime}$
and is very bright in the first season. Together, these variables
cause the i$^{\prime}$ baseline to be be rather messy.

\noindent{\it Event 7}: The baselines of this very high S/N event are
quite well behaved in both r$^{\prime}$ and i$^{\prime}$, but there is
a quite bright variable source located at 15 pixels, which may cause
some deviations. However, the deviations in the i$^{\prime}$ data from
the Paczynski fit between days 100 and 150 are too large to be caused
by these near neighbours. Also from this event, some thumbnails are
shown in figure \ref{fig:thumbs}.

\noindent{\it Event 8}: Even though there is a variable source located
at 7 pixels distance from this event, the baselines are quite well behaved.

\noindent{\it Event 9}: Two faint variables are located at about 4 and 8 pixels
distance from this event, and a bright one at 12 pixels. The
lightcurves do not appear to be affected substantially.

\noindent{\it Event 10}: Here the noisy baseline in i$^{\prime}$ and the two deviating points
in the r$^{\prime}$ lightcurve around day 100 can be attributed to the same
variable source 7 pixels away.

\noindent{\it Event 11}: While the r$^{\prime}$ baseline of this high
amplification event is 
well behaved, the i$^{\prime}$ baseline is not completely flat. In the
difference images there is a hint of a weak variable source a few
pixels away, which is probably the cause of the small baseline wiggle.
See figure \ref{fig:thumbs} for thumbnails of this event.

\noindent{\it Event 12}: The r$^{\prime}$ lightcurve of this event is quite well behaved and
there are no nearby variable sources. The reason for the noise in the
i$^{\prime}$ lightcurve is not clear.

\noindent{\it Event 13}: A fairly bright variable source is situated 16 pixels away
from this event, but the lightcurves in both r$^{\prime}$ and i$^{\prime}$ are well
behaved.

\noindent{\it Event 14}: The noise in the lightcurves and more particularly the small
bump in the i$^{\prime}$ lightcurve around day 370 is due to a variable source 3
pixels away from the event.

\subsection{POINT-AGAPE events}

\begin{figure}
\includegraphics[width=8cm]{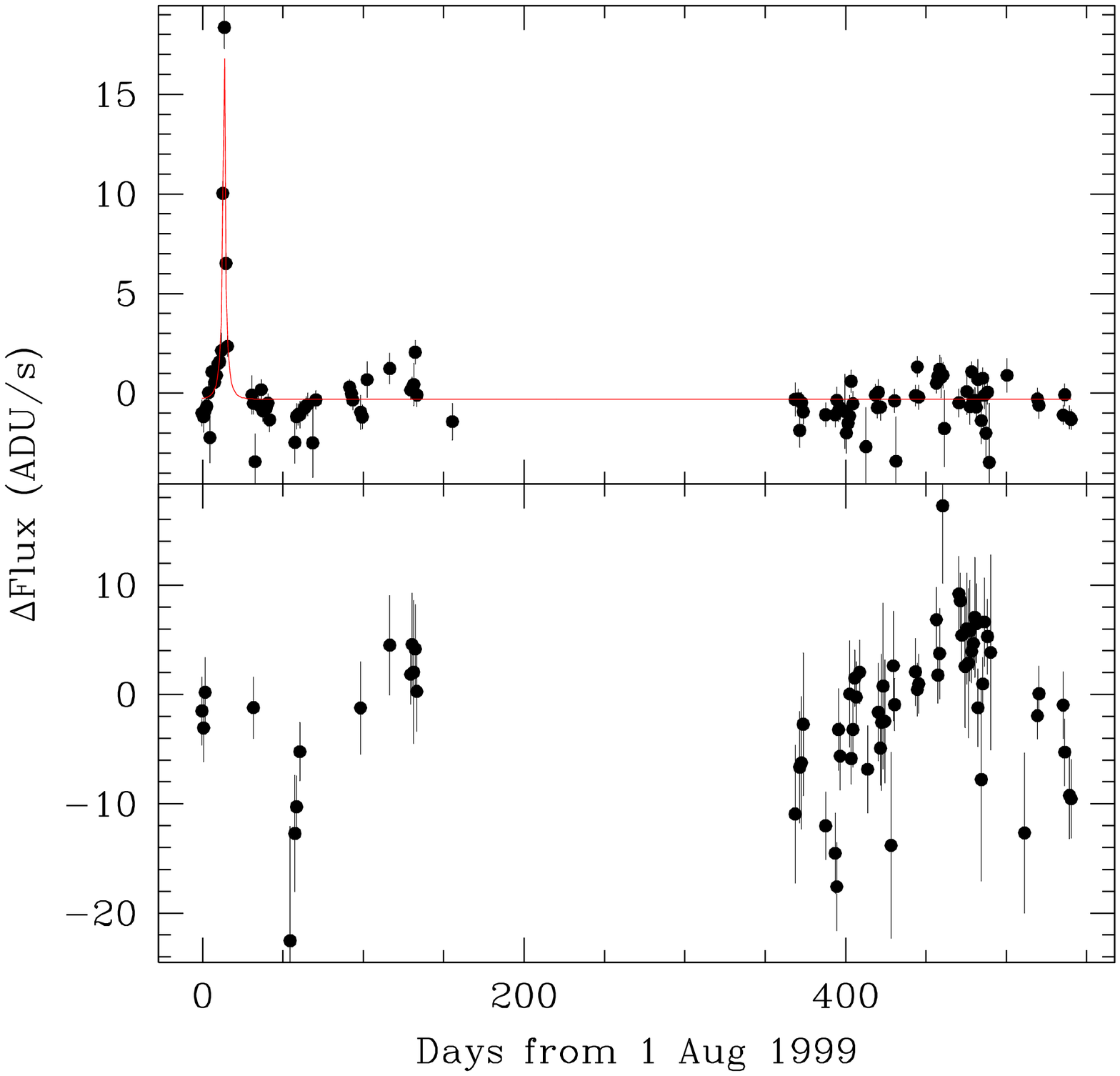}
\caption{The lightcurves in r$^{\prime}$ (upper panel) and i$^{\prime}$ (lower panel) of
the microlensing event PA-99-N1, reported by POINT-AGAPE. In the i$^{\prime}$
data the variability in the baseline,
caused by nearby variables, is high enough to give a $\chi^2$ which is
too high to pass our selection procedure.}
\label{fig:agapen1}
\end{figure}

The POINT-AGAPE group uses the same data for their microlensing survey
towards M31, although their pixel lensing technique is different
(Paulin-Henriksson et al. \cite{Paulin03}). Based on the same two-year
dataset, they find four high signal-to-noise microlensing events. Two
of these events are also present in our sample, namely PA-99-N2, our
event number 7, and PA-00-S4, our event number 11.

The fact that our sample of candidate microlensing events is much
larger is caused by the severe selection criteria used by POINT-AGAPE,
that limit the number of events found. For example, POINT-AGAPE demands
FWHM timescales smaller than 25 days and much higher signal-to-noise
than we.
The following two events reported by the POINT-AGAPE team are not in our
sample of candidate microlensing events.\\
$\bullet$ {\it PA-99-N1}\\
This event did not pass our filtering procedure because of too large
 variability in the i$^{\prime}$ baseline caused by a very closeby variable
star. Our lightcurves of this event are shown in figure
\ref{fig:agapen1}.\\
$\bullet$ {\it PA-00-S3}\\
Very close to the bulge of M31 the surface brightness becomes very
high and our difference images become very noisy and of low
quality. This event is located in a part of the field that is not used
for the current analysis for this reason. In figure \ref{fig:candpos}
the region that is not used in our analysis is indicated.

\section{Discussion}

Due to the nature of M31 microlensing, it is very difficult to determine
the nature of the lensing object in individual microlensing
events. In classical microlensing the mass of the lens, the relative
motion of the lens with respect to the source and the distance to the
lens are usually unknown and degenerate parameters.
In difference-image lensing there is an additional unknown, namely the
unlensed flux of the source.
If a conclusion about the existence and/or nature of a MACHO
population in the halo of M31 is to be drawn from a microlensing
survey, statistical methods have to be used.

By modeling the population of stars that can be lensed and the
population of objects that can act as lenses, predictions can be made
about the rate of occurrence, the spatial distributions, and the
timescales of microlensing events. All these data can be used to
constrain the lens populations responsible for observed microlensing
events (Baltz, Gyuk \& Crotts \cite{Baltz03}, hereafter BGC; Kerins et
al. \cite{Kerins01}).

To compare our sample of candidate microlensing events with rate maps
as published by BGC, the detection efficiencies and observing
characteristics have to be carefully modeled, which will be done in a
future paper. Instead of a full statistical analysis of the current
sample, we restrict our discussion to a simple analysis of the spatial
and timescale distributions of the candidate events to see if they are
consistent with predictions for halo lensing and lensing by stars in M31.

\subsection{Timescale distribution}

\begin{figure}
\includegraphics[width=8cm]{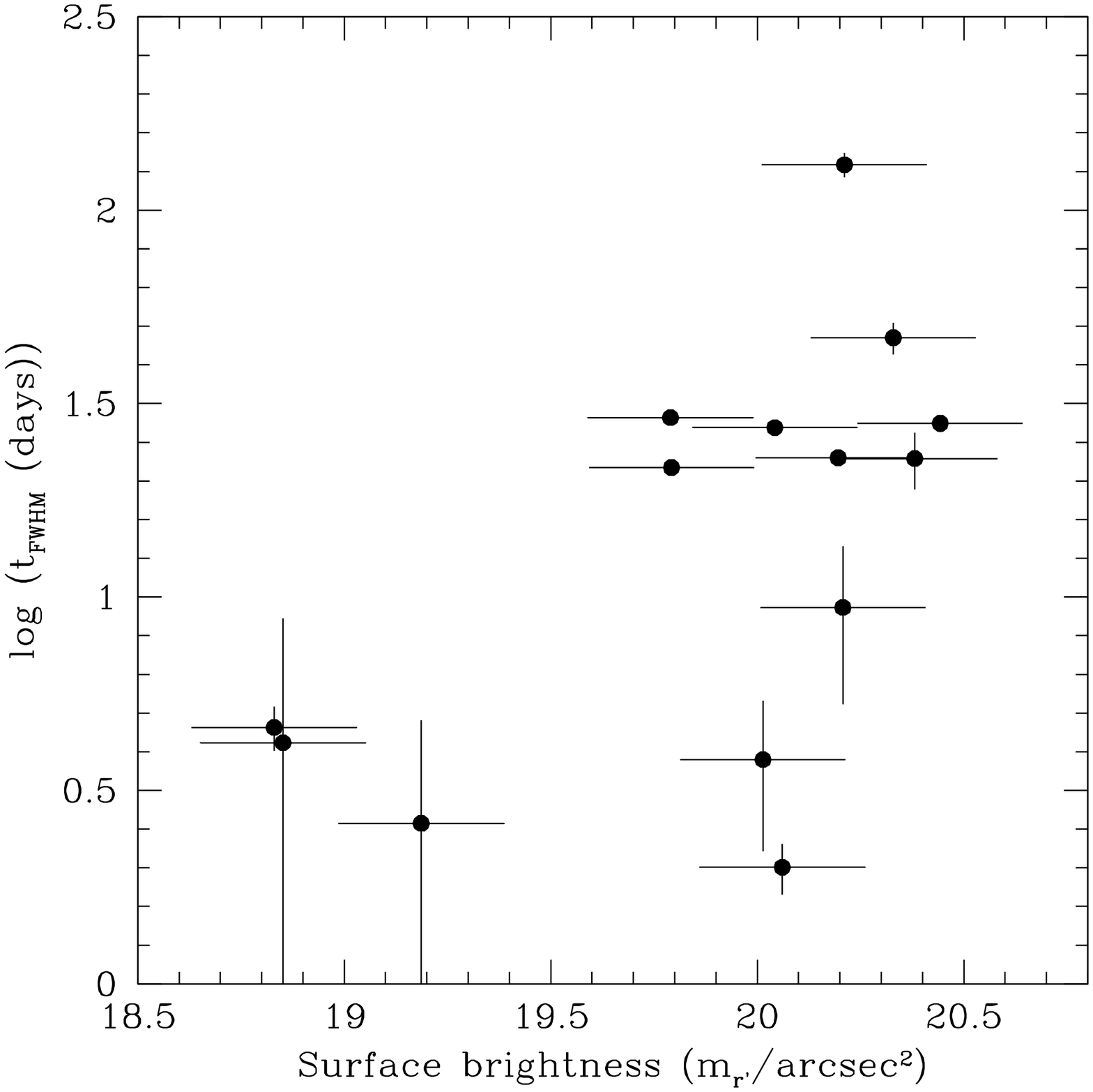}
\caption{For the 14 candidate microlensing events, the logarithms of
the FWHM timescales are plotted against the local r$^{\prime}$
surface brightnesses. At high surface brightnesses the timescales tend
to be shorter than in low surface brightness regions.}
\label{fig:tfvssb}
\end{figure}

The timescale of a microlensing event depends on the relative
velocities of the source and the lens and the size of the Einstein
radius, which in turn depends on the mass of the lens and the geometry
(eq. \ref{eq:r_e}). In the case of M31 microlensing we can assume that
$D_{OL} ~=~ D_{OS}$ and since the Einstein ring crossing time $t_E ~=~
2R_E/v_{\perp}$ we get:
\begin{equation}
t_E \propto M_L^{1/2} D_{LS}^{1/2} v_{\perp}^{-1}
\end{equation}
where $v_{perp}$ is the relative speed between lens and source
projected on the sky.
The $t_{FWHM}$ that we measure is related to $t_E$ according to
equation \ref{eq:tfwhm}.

For self-lensing, i.e. lensing by stars, this translates into a
timescale distribution that varies spatially. In the bulge, velocities
are relatively high and distances between lens and source relatively
short (compared to lensing by halo objects), therefore event durations
will be relatively short. In the disk, relative velocities are much
slower and event durations longer than in the bulge. On top of that,
the timescales are expected to rise when going to larger radii, since
the random velocities of the stars in a galactic disk are proportional
to the root of the surface density (Bottema \cite{Bottema93}).
Roughly speaking we can say that average self-lensing timescales are
inversely correlated with surface brightness; in higher surface
brightness regions the timescales are expected to be shorter.

For halo lensing, relative velocities are higher than for disk-disk
lensing, but because of the larger lens-source distance the event
durations are still relatively long. Because the relative velocities
and distances do not change much within the observed field, the
timescale distribution of halo-lensing events should be spatially
uniform.  Plots of the expected timescale distributions for different
lines of sight towards M31 are given by BGC for stellar self-lensing
and halo lensing combined and reflect these trends.

In figure \ref{fig:tfvssb} we plot for our 14 microlensing candidates the
logarithm of $t_{FWHM}$ versus the r$^{\prime}$ surface brightness at their
locations. The candidate events close to the center of M31, in
locations with surface brightness brighter than 19.5 
$mag_{r\prime}\, arcsec^{-2}$,
all have event durations in the order of a few days. Most of the
candidates in the low surface brightness regions have significantly
longer timescales. That this distribution is exactly what is expected for
microlensing and the fact that the timescales are consistent with the
timescales predicted by BGC are further indications that our sample of
candidates consists of actual microlensing events.

\subsection{Spatial distribution}

\begin{figure*}
\includegraphics[width=16cm,clip=]{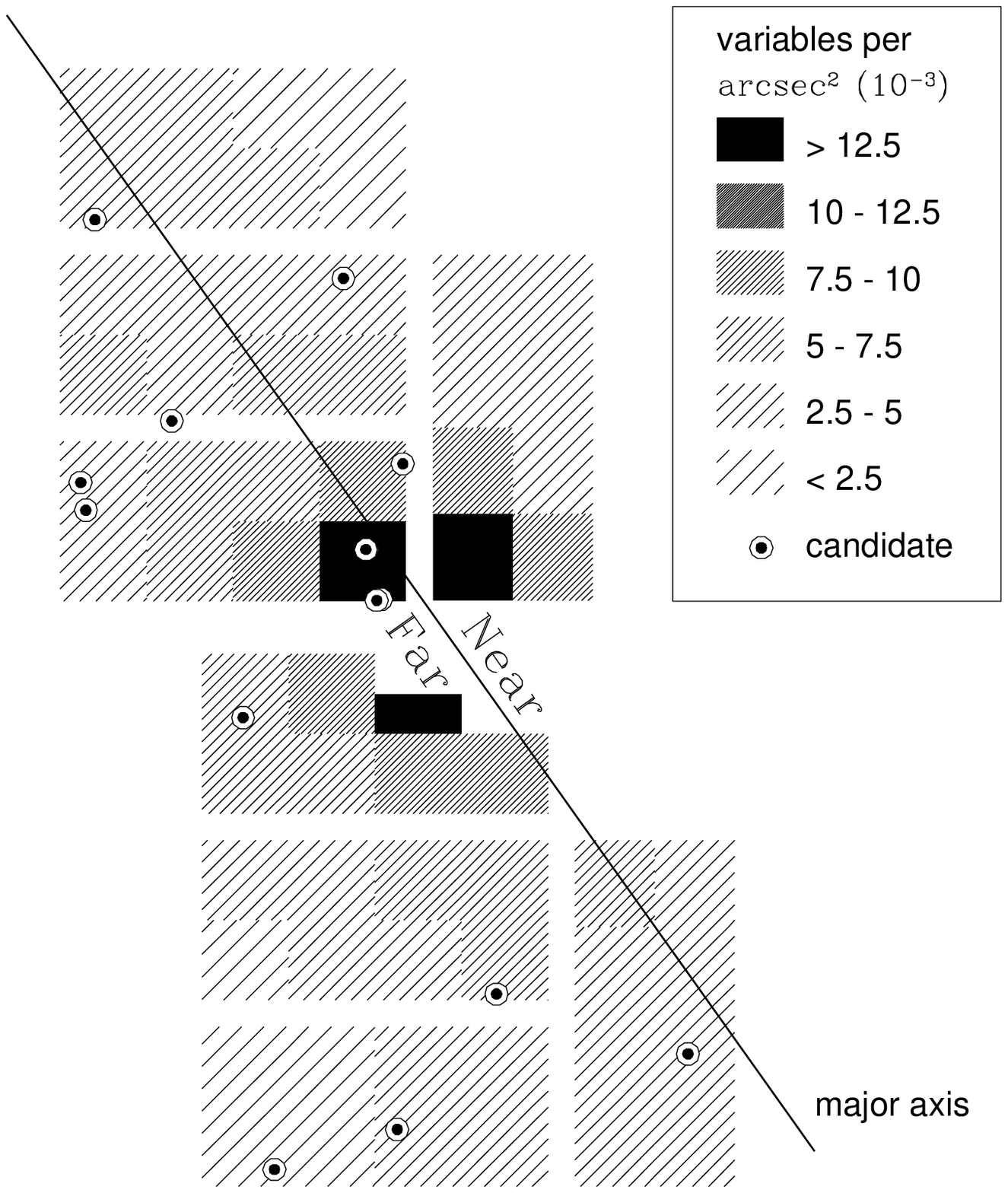}
\caption{Surface density map of long period variable stars in the
surveyed area. Each chip was subdivided in 2x4 rectangles
(300x324\arcsec), in which the number of variables with periods longer
between 150 and 600 days was counted. Only stars with accurately
determined periods were used and the edges of the chips were avoided
to ensure only well sampled lightcurves were used, leaving 32,841
variables. The positions of the 14 microlensing candidates are
indicated, as well as the major axis of M31.}
\label{fig:vardistr}
\end{figure*}

The expected spatial distribution of stellar self-lensing events is
strongly concentrated to the center because both the density of
possible source stars and lenses is highest there. Furthermore, the
spatial distribution is nearly symmetric from the far to the near side
of the disk of M31 (BGC), so in the case of only self-lensing the
number density of events should be the same at the same radius on both
the far and near side. The stellar halo of M31 has not been taken into
account by BGC, but is not expected to change the symmetry
significantly, since, like the bulge, it acts both as a lens and as a
source population. If the spheroid has the same M/L as the disk, for
every event caused by a spheroid-lens and disk-source pair, there is a
disk-lens and spheroid-source pair event, as is the case with
disk-bulge and bulge-disk lensing (BGC). In case of heavy extinction
in the disk, the spheriod could induce an asymmetry in the lensing
rate.

Dark halo lensing on the other hand will induce an asymmetric microlensing
distribution, because it only acts as a lens population. Due to the
longer line-of-sight through the M31 halo towards the far side of the
disk, the microlensing optical depth is much higher on the far side
than on the near side. Therefore, an asymmetry in the density of
microlensing events between the far and near side is a strong
indication of a microlensing halo. Halo lensing is also more likely
near the center because of the higher surface density of possible
source stars, but drops off much more slowly, since the surface
density of halo lenses should not drop off very quickly.

Without knowing the detection efficiency of microlensing events, the
number of detected candidate events is hard to interpret
quantitavely. Since the detection efficiency might vary with position,
their spatial distribution is also hard to interpret.  However, if we
assume that the spatial distribution of variable stars is symmetric
over M31, we can get an idea of the variation of the detection
efficiency with position by looking at the spatial distribution of the
detected variable stars.  The variable stars that are most likely to
be mistaken for microlensing are variables with periods longer than
150 days.  In table \ref{tab:miras} we present the detected numbers of
variables with periods between 150 and 600 days, for
300x324$^{\prime\prime}$ subregions in our field of view. For each
subregion the actual area, corrected for area lost due to bright stars
and diffraction spikes, is also tabulated. In figure
\ref{fig:vardistr} we show the corresponding variable star density
map, with the positions of the 14 candidate events plotted on top.

To select the long-period variables (LPVs) from the complete set of
lightcurves, all lightcurves were checked for periodic behaviour using
the Numerical Recipes (Press et al. \cite{numrec}) algorithm based on
the Lomb method (\cite{Lomb}) for spectral analysis of unevenly
sampled data. The lightcurves for which a 
periodicity between 150 and 600 days was found with very high
significance were selected. In total 32,841 variables were selected,
excluding the part of the south field close to the bulge and one other
small part at the edge of the south field, where a very bright
foreground star is located.

In the spatial distribution of the LPVs we see
no surprises. The surface density of these objects is correlated with
the surface brightness. 
Since we are interested in a possible asymmetry in the
distribution of the candidate microlensing events, we compare the
relative distributions of variables and events over the near and the
far side of M31. We assume that the LPVs are a good tracer of the
population of possible microlensing source stars and that the
chances of detecting LPVs and microlensing events are the same. The
detection efficiencies of both will depend on the surface brightness
of the disk and be subject to extinction effects. 

\begin{table*}
\centering
\begin{tabular}{llcc|llcc}
\hline
\hline
RA & DEC & LPV & Area & RA & DEC & LPV & Area\\
(J2000) & (J2000) & \# & ($\Box^{\prime}$) & (J2000) & (J2000) & \# & ($\Box^{\prime}$)\\
\hline
0:44:43.45 & 41:24:49.3 & 444 & 26.00 & 0:42:13.67 & 41:36:13.5 & 241 & 26.18\\
0:44:42.78 & 41:19:51.5 & 390 & 25.91 & 0:42:40.14 & 41:36:14.6 & 269 & 26.25\\
0:44:14.82 & 41:24:55.5 & 577 & 26.37 & 0:42:13.68 & 41:30:50.9 & 248 & 26.10\\
0:44:14.21 & 41:19:57.3 & 534 & 26.27 & 0:42:40.13 & 41:30:51.2 & 349 & 26.08\\
0:43:46.07 & 41:25:1.6 & 666 & 25.92 & 0:42:13.77 & 41:25:28.0 & 401 & 26.02\\
0:43:45.53 & 41:20:3.3 & 754 & 26.38 & 0:42:40.19 & 41:25:27.6 & 739 & 26.25\\
0:43:17.27 & 41:25:7.9 & 887 & 26.16 & 0:42:13.92 & 41:20:6.0 & 761 & 25.92\\
0:43:16.80 & 41:20:9.4 & 1340 & 26.36 & 0:42:40.32 & 41:20:5.0 & 1428 & 24.34\\
0:44:44.58 & 41:47:58.6 & 369 & 26.47 & 0:44:44.39 & 41:36:30.8 & 422 & 26.21\\
0:44:44.15 & 41:43:4.1 & 446 & 26.42 & 0:44:43.88 & 41:31:32.6 & 463 & 25.64\\
0:44:15.93 & 41:48:8.2 & 253 & 26.26 & 0:44:15.69 & 41:36:38.1 & 317 & 25.57\\
0:44:15.46 & 41:43:13.1 & 378 & 26.51 & 0:44:15.20 & 41:31:39.4 & 436 & 26.14\\
0:43:47.16 & 41:48:16.3 & 202 & 26.40 & 0:43:46.87 & 41:36:44.7 & 412 & 25.60\\
0:43:46.65 & 41:43:20.9 & 244 & 25.38 & 0:43:46.39 & 41:31:45.6 & 519 & 26.03\\
0:43:18.32 & 41:48:23.1 & 239 & 26.64 & 0:43:18.00 & 41:36:50.4 & 351 & 26.15\\
0:43:17.78 & 41:43:27.5 & 186 & 25.53 & 0:43:17.53 & 41:31:51.3 & 599 & 25.97\\
0:43:52.75 & 40:48:54.8 & 139 & 26.78 & 0:41:24.61 & 41:0:17.3 & 383 & 26.24 \\
0:43:52.20 & 40:43:57.2 & 100 & 26.61 & 0:41:50.85 & 41:0:18.2 & 480 & 25.75\\
0:43:24.36 & 40:48:59.6 & 203 & 26.52 & 0:41:24.63 & 40:54:54.3 & 375 & 26.35\\
0:43:23.88 & 40:44:1.4 & 155 & 26.54 & 0:41:50.83 & 40:54:54.9 & 390 & 25.41\\
0:42:55.86 & 40:49:4.3 & 362 & 26.57 & 0:41:24.71 & 40:49:31.4 & 357 & 26.36\\
0:42:55.46 & 40:44:5.9 & 246 & 26.58 & 0:41:50.86 & 40:49:31.5 & 402 & 25.67\\
0:42:27.33 & 40:49:8.9 & 421 & 26.65 & 0:41:24.88 & 40:44:9.4 & 393 & 26.53\\
0:42:26.99 & 40:44:10.5 & 263 & 26.53 & 0:41:50.97 & 40:44:8.8 & 425 & 25.41\\
0:43:52.75 & 40:48:54.8 & 550 & 26.01 & 0:43:53.25 & 41:0:37.5 & 284 & 26.30\\
0:43:52.20 & 40:43:57.2 & 492 & 26.42 & 0:43:52.89 & 40:55:39.1 & 202 & 26.07\\
0:43:24.36 & 40:48:59.6 & 877 & 26.05 & 0:43:24.79 & 41:0:43.1 & 406 & 25.57\\
0:43:23.88 & 40:44:1.4 & 594 & 25.80 & 0:43:24.46 & 40:55:44.3 & 277 & 26.22\\
0:42:55.76 & 40:47:49.6 & 690 & 13.16 & 0:42:56.21 & 41:0:47.7 & 540 & 25.43\\
0:42:55.46 & 40:44:5.9 & 799 & 25.49 & 0:42:55.89 & 40:55:48.7 & 474 & 26.33\\
0:42:26.99 & 40:44:10.5 & 847 & 25.06 & 0:42:27.61 & 41:0:51.6 & 568 & 26.45\\
~& ~& ~& ~& 0:42:27.29 & 40:55:52.4 & 564 & 25.73\\
\hline
\end{tabular}
\begin{small}
\caption {Number counts of detected long-period variable (LPV)
stars. For each subregion, the coordinates of the center, the number
of LPV's and the area, corrected for area lost to bright stars and
spikes, are given.}
\label{tab:miras}
\end{small}
\end{table*}

\begin{figure}
\includegraphics[width=8cm]{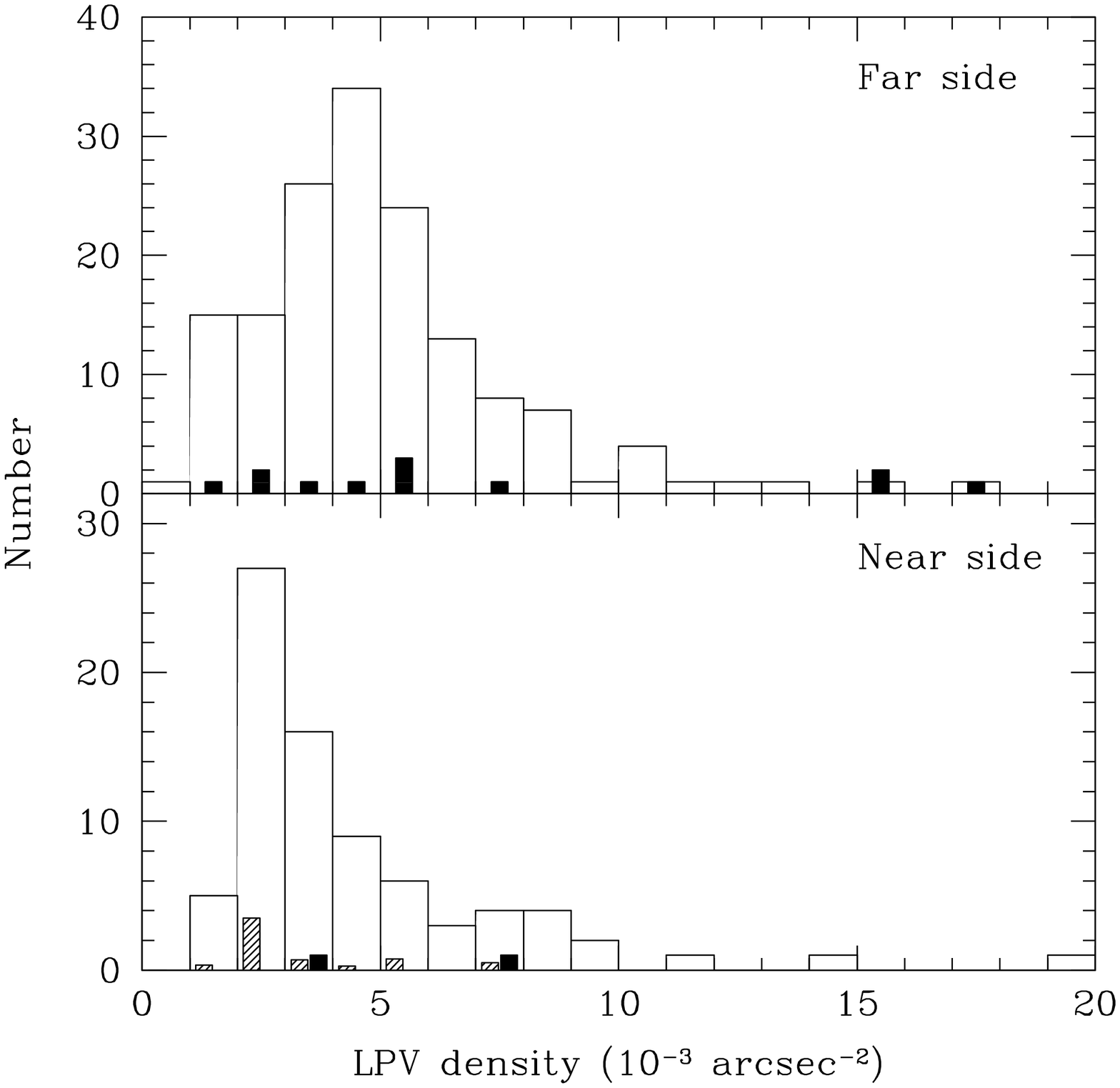}
\caption{Upper panel: histogram of the surface densities of
long-period variable stars (LPVs) in small
150x162$^{\prime\prime}$ regions on the far side of the M31 disk. The
thin, black bars show in which LPV surface density bins the
far side microlensing candidate events are located. Lower panel:
histogram of the surface densities of LPVs in small
150x162$^{\prime\prime}$ regions on the near side. The thin, shaded
bars show the numbers of candidate microlensing events predicted
by scaling the numbers of far side events with the ratio of regions on
the near and far side in each LPV density bin. In total 6 events are
expected on the near side, but only 2 are detected, shown by the thin,
black bars.}
\label{fig:lpvhist}
\end{figure}

To investigate the different parts of the far and near side of the M31
disk that we sample, we divide each subregion in table \ref{tab:miras}
in 4 smaller 150x162$^{\prime\prime}$ pieces. In the upper panel
of figure \ref{fig:lpvhist} we plot the distribution of the LPV
surface densities of these small subregions. The smaller, black
histogram shows the distribution of the far side candidate events over
the LPV density bins. 
If the microlensing rates are intrinsically near/far side symmetric,
as we assume the LPVs to be, the expected number of candidate events
on the near side can be determined for each LPV density bin, by
scaling the number of far side events. We can only do this for the low
surface density bins, since the high surface density bins are not
populated sufficiently. However, since close to the center of M31 the
microlensing optical depth is dominated by stellar self-lensing they
should not affect the symmetry of the spatial distribution.
The lower panel of figure \ref{fig:lpvhist} shows the distribution of
the subregions on the near side over the LPV density bins. The shaded
histogram shows the distribution of the far side candidate events
scaled to the near side LPV density distribution, and the black
histogram the actual near side candidate events.

Whereas 6 near side candidate events are predicted from the distribution
of the 9 far side events in areas with fewer than 10 LPVs per square arcsec, only 2 are detected. 
Thus, the distribution of candidate events seems to be more
near/far asymmetric than the distribution of variable stars, which is
an indication that it is a separate population of objects and not some
kind of leakage from the population of LPVs. As argued
before, the most likely explanation of such an asymmetry of
microlensing events is dark halo lensing. The number of
candidate events is still quite small, but the chance of finding a
far:near split as strong as 9:2 is only 12\%. 
It has been argued by Paulin-Henriksson et al. (\cite{Paulin02}) that
candidate event 11 is possibly caused by a stellar lens in M32 and not
by a lens in the halo. Doing the same calculation without candidate
event 11, the chance of finding the 8:2 ratio is 16\%.

\section{Conclusions}

In this paper we present the first 14 microlensing candidates from the
MEGA survey and a preliminary discussion of their properties.  Our
photometry turns out to be sensitive to nearby variable sources in a
way that causes us to miss out on events like PA-99-N1. It also leads
to the necessity of a by-eye selection step in the filtering of the
lightcurve database.  Before the photometry is improved, enabling the
filtering procedure to be automated and a detailed efficiency
calculation to be done, we cannot draw firm conclusions based on the
spatial distribution of our sample of candidate microlensing
events. However, the spatial distribution of our candidate
microlensing events is more near:far asymmetric, than would be
expected from the spatial distribution of the detected long-period
variable stars. Although the significance of this asymmetry is low,
due to the small number of events and the lack of a detailed detection
efficiency, the spatial distribution of the candidate microlensing
events is suggestive of the presence of a microlensing halo.

Thus far we have only extracted candidate microlensing events from
the 1999/00 and 2000/01 seasons. As a next
step the remaining seasons will also be searched for microlensing
events. Since we also observe with other telescopes than the INT, much
more events will likely be detected when the data from the different
telescopes are combined. Especially the sensitivity to short timescale
events will benefit from merging the INT data with the data obtained
with the telescopes of the MDM observatory, as they are supplementary
in time coverage. Fainter events will also become easier to detect because
the MDM data go deeper than the INT data, as do the data obtained with
the KPNO 4m telescope.

\begin{acknowledgements}
We would like to thank all observers who have performed the
observations for this survey. We also wish to thank the referee for
his helpful comments. A.C. was supported by grants from NSF (AST
00-70882, 98-02984, 95-29273 and INT 96-03405) and STScI (GO 7376 and
7970).\\
\end{acknowledgements}

\appendix

\section{Microlensing event selection} 

A microlensing event
caused by a single lens has a characteristic shape and a flat
baseline. Such a standard microlensing lightcurve is called a
Paczynski lightcurve (Paczynski \cite{Pacz86}) and is described by
\begin{equation}
F(t) ~=~ F_0 \times A(t) ~=~ F_0 \times \frac{u^2 + 2}{u \sqrt{u^2 + 4}}
\label{eq:fluxoft}
\end{equation}
where $F_0$ is the baseline, unlensed flux, $A(t)$ is the
amplification, and $u$ is the projected
distance between the lens and the source, in units of the Einstein
radius. 
This Einstein radius depends on the geometry of the system and the
mass of the lens and in the lens plane is given by:
\begin{equation}
R_E = \sqrt{\frac{4Gm}{c^2} \frac{D_{OL}D_{LS}}{D_{OS}}}
\label{eq:r_e}
\end{equation}
where $m$ is the lens mass and the Ds are the distances between observer,
lens and source.
Note that the amplification is always 1 or higher and independent of
wavelength, meaning that microlensing is in principle achromatic and conserves the
colour of the source. If the relative motion of lens and source is
taken to be a uniform motion, then $u$ can also be written as:
\begin{equation}
u(t) = \sqrt{u_{min}^2 + \left( \frac{t-t_{max}}{t_E}\right)^2 }
\end{equation}
where $u_{min}$ is the impact parameter, $t_{max}$ the time of maximum
amplification and $t_E$ the Einstein time. This is defined as the time it
would take the source to cross the Einstein radius.
Since we are measuring only the flux difference of variable objects
with respect to a template image, equation \ref{eq:fluxoft} transforms
into:
\begin{equation}
\Delta F(t) ~=~ \Delta F_{bl} + F_0 \times (A(t)-1)
\label{eq:dfluxoft}
\end{equation}
where $\Delta F_{bl}$ is the baseline flux minus the flux on the
reference image. As the crowding in M31 prevents the unlensed source
flux from being measured, $\Delta F_{bl}$ is an unknown parameter that
has to be fitted.

\begin{figure}
\centering
\includegraphics[width=8cm]{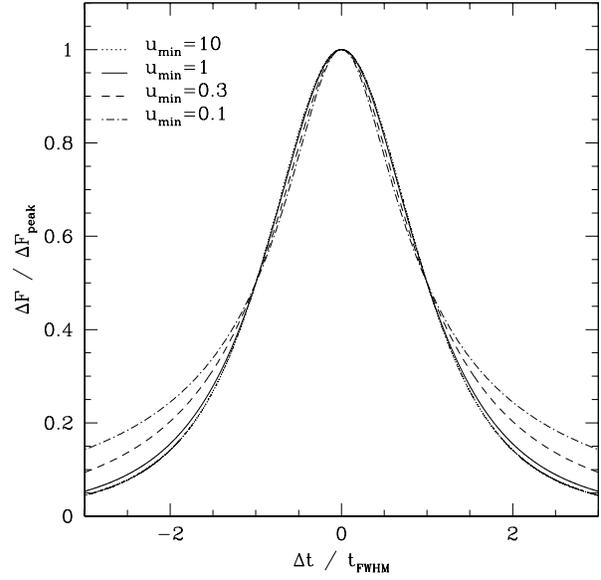}
\caption{Differences in the Paczynski lightcurves for different values
of the impact parameter $u_{min}$. The peak shape does not depend
strongly on $u_{min}$, except from in the wings. In M31,
where microlensing events are often only resolved while significantly
magnified, the wings are usually not strongly constrained, meaning
that it is difficult to measure $u_{min}$.}
\label{fig:pacshapes}
\end{figure}

Furthermore, since the baseline flux is unknown and the microlensing
event resolved only while magnified, it is very difficult to measure
$t_E$ and $u_{min}$ (Gould \cite{Gould96}; Baltz \& Silk \cite{BaltzSilk00}), consequently these two parameters are highly
degenerate, as shown in Figure \ref{fig:pacshapes}. Instead, we fit
the width of the peak at half of the maximum flux 
$t_{FWHM}$ (Gondolo \cite{Gondolo99}), which is related to $t_E$ by:
\begin{eqnarray}
\label{eq:tfwhm}
& t_{FWHM} = t_E \cdot w(u_{min})\\
\textrm{where} & w(u_{min}) = 2 \sqrt{2 f( f(u_{min}^2)) - u_{min}^2 }\\
\textrm{with} & f(x) = \frac{x + 2}{\sqrt{x(x + 4)}} - 1
\end{eqnarray}
We perform a four-parameter fit by letting $t_{FWHM}$ replace $t_E$ and
$u_{min}$. To save computer time we assume a fixed value for $u_{min}$ and let
$t_E$ float to fit for $t_{FWHM}$. Since there is some difference in
the shape of the peak depending on $u_{min}$, we make fits for three
values of $u_{min}$, namely 0.1, 0.3 and 1.0. These fits are
sufficient for our selection procedure. Figure
\ref{fig:pacshapes} shows that this range covers the range of possible
shapes very well.

\subsection{Selection of r$^{\prime}$ lightcurves}

The selection of microlensing candidates was based on the goodness of
fit of the standard Paczynski lightcurve fits to the 99/00 and 00/01 data.
Based on the parameters of the best fits, the lightcurves are
filtered in several steps, described below.\\
$\bullet$ {\it Peak sampling}\\
To be able to constrain event parameters it is important to sample the
peak of the lightcurve well. We demand that at least the part of the
peak where the flux is more than half of the peak flux lies completely
within an observing season. For this the $t_{max}$ and $t_{FWHM}$ from the
Paczynski fits are used. We also demand that there are at least 2 data
points above 25\% of the fitted peak flux. Furthermore, we calculate
a `peak weight' $P$ defined as the sum of all data points within the peak:
\begin{equation}
P ~=~ \sum_{i \in peak} \frac{\Delta F_i - \Delta F_{bl}}{\sigma_i}
\end{equation}
where $\Delta F_i$ is the difference flux, $\sigma_i$ the error on the
difference flux, and $\Delta F_{bl}$ is the difference flux of the
baseline, as given by the Paczynski fit. All the points that are
within 2 $t_{FWHM}$ of $t_{max}$ are considered to be in the peak
and are used for this calculation. All points farther away from the
center of the peak are considered to be part of the baseline.
A minimum `peak weight' $P$ of 25 is demanded.\\
$\bullet$ {\it Peak significance}\\
To exclude spurious detections, a minimum peak significance is
demanded. The $\chi^2$ value of a flat line fit to the
lightcurve is compared to the $\chi^2$ of the Paczynski fit. The
difference between these two values has to be larger than 100 for
selection.\\
$\bullet$ {\it Peak width}\\
Both a minimum and a maximum peak width are used. The minimum
$t_{FWHM}$ of 1 day serves as an extra filter against spurious
detections. Events with FWHM timescales longer than 150 days are also
excluded, since with the current dataset the baseline is not well
sampled. In practice, fits that give these very long timescales are
usually caused by lightcurves that continue to rise at the end of an
observing season.\\
$\bullet$ {\it Baseline flatness}\\
Microlensing events have flat baselines, contrary to periodic
variables. The flatness of the baseline is checked by the goodness of
fit of the Paczynski curve to the baseline, where the baseline is
defined as that part of the lightcurve that is more than twice the
$t_{FWHM}$ away from $t_{max}$. For this part of the lightcurve we use a cut of
$\chi^2_{bl}/N < 1.5$, where $N$ is the number of points in the
baseline. The fits to the ``empty'' lightcurves that were used to
assess the quality of the difference images showed that such a rather
lenient cut is necessary, because our current photometry turns out to
be quite sensitive to nearby variable objects which can cause some
additional variability in some lightcurves.\\
$\bullet$ {\it Goodness of fit}\\
Finally, the shape of the peak must be consistent with microlensing,
meaning that the Paczynski function must give a good fit. For the
Monte Carlo simulations we find that a $\chi^2/N$ cut of 1.2 includes
90\% of all events. However, secondary effects, like for example
parallax effects, can influence the exact shape of the
lightcurve. Also, the possibility of additional variability due to
nearby variable objects must be taken into account.  We use a rather
lenient $\chi^2/N$ cut again, and since secondary effects will be
stronger in high signal-to-noise events, this cut is also dependent on
the `peak weight' $P$ of the peak. The cut is described by
$\chi^2_{pk}/N_{pk} < 1.5 + (\frac{P}{N_{pk}}-1) \times 0.1$, where
$N_{pk}$ is the number of points in the peak, as defined
above. Formally, a $\chi^2/N$ cut of 1.5 corresponds to a
probability of 0.1\% for the degrees-of-freedom in our fits,
meaning that 99.9\% of perfect microlensing lightcurves without any 
additional variability would pass this criterium.

\begin{figure}
\includegraphics[width=8cm]{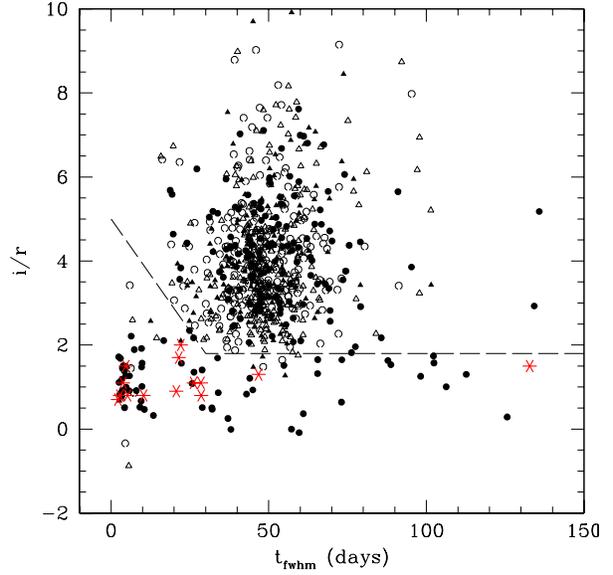}
\caption{Plot of the peak width and colour of the peaks that pass our
r$^{\prime}$ filtering procedure. The different symbols refer to the $\chi^2/N$
value of the Paczynski fits to the i$^{\prime}$ lightcurves. Closed circles are
$\chi^2/N < 2$, open circles $2 < \chi^2/N < 3.5$, closed triangles $3.5 <
\chi^2/N < 5$, and open triangles are $\chi^2/N > 5$. The positions of the
14 candidate microlensing events are indicated with stars. Clearly
visible is that the long period variable stars cluster around a
$t_{FWHM}$ of 50 days and a i$^{\prime}$/r$^{\prime}$ flux of 4. For further filtering,
the colour cut indicated by the dashed line was used. A $\chi^2/N$ of 2
was used as the base value of the $\chi^2/N$ cut applied to the
Paczynski fits to the i$^{\prime}$ data. The remaining sample still contains
lightcurves with variable baselines and events caused by bad pixels.}
\label{fig:tfcolour}
\end{figure}

\subsection{Colour information}

For several reasons the i$^{\prime}$ band data are important for the
microlensing candidate selection. First, gravitational lensing is
achromatic, so that the colour of the observed light in the difference
images will not
change during the microlensing event. Most periodic variable
stars on the other hand, change colour during their pulsation
cycles. Second, the long period variable stars like Mira's that can
otherwise easily be mistaken for microlensing events, are very cool
and much brighter in i$^{\prime}$ than in r$^{\prime}$. Also the
variability is much larger and therefore easier to see in i$^{\prime}$
than in r$^{\prime}$. Often, baselines that look flat in the
r$^{\prime}$ data, clearly show variability in i$^{\prime}$. Third,
most long period variables turn out to have similar colours, providing
an easy way to filter them out of the lightcurve database.

Paczynski fits are done to the two-season i$^{\prime}$ lightcurves of
the microlensing candidates selected from the r$^{\prime}$ data. The
values for the wavelength-independent lightcurve parameters $t_{max}$,
$t_{FWHM}$ and $u_{min}$ are taken from the fits to the
r$^{\prime}$ data and used in the i$^{\prime}$ fits, since the
r$^{\prime}$ data constrain these parameters better because of their
superior quality. In the case of large colour changes during the
event, these Paczynski fits to the i$^{\prime}$ lightcurves should be
poor, indicative of a non-microlensing event.\\
$\bullet$ {\it Colour cut}\\
As mentioned before, long period variable stars have cool atmospheres
and therefore are very red. In figure \ref{fig:tfcolour}, the colours
of the selected transients are shown as a function of the $t_{FWHM}$
of the Paczynski fits. The long period variable stars clearly stand
out and cluster around a $t_{FWHM}$ of 50 and an
i$^{\prime}$/r$^{\prime}$ flux ratio of 4. To remove the long period
variables from the sample, a colour/$t_{fwhm}$ cut is used, indicated
by the dashed line.\\
$\bullet$ {\it Goodness of fit}\\
In the i$^{\prime}$ data, the
problems of crowding of the residuals in the difference images is 
worse than in the r$^{\prime}$ data, meaning that our photometry is
influenced by nearby variables even more than in the r$^{\prime}$ data. Since
the long period variables are very red, more are detected in the i$^{\prime}$
difference images and the residuals are much brighter. In figure
\ref{fig:tfcolour} the transients are divided into four
different bins depending on the $\chi^2/N$ value of the Paczynski fit to
the i$^{\prime}$ data. A similar `peak weight' dependent $\chi^2/N$ cut
was applied to the i$^{\prime}$ lightcurves as to the peak of the r$^{\prime}$ lightcurve, but with a minimum $\chi^2/N$ of 2 in
stead of 1.5: $\chi^2/N < 2.0 + (\frac{P}{N_{pk}}-1) \times 0.1$.

\section{Monte Carlo simulations}

In order to develop and test the efficiencies of our lightcurve
filtering procedures to select
microlensing lightcurves from a large set of lightcurves, we performed
Monte Carlo simulations of microlensing events. To be meaningful for
comparison with the data, the simulated lightcurves should have the 
same time sampling and error distribution as the
data. Furthermore, a broad range of time scales, times of maximum
amplification, and peak brightnesses has to be sampled.

In total 170,000 lightcurves were simulated with random peak times,
$t_{FWHM}$ of 1, 3, 6, 10, 30 or 60 days, seven different peak fluxes,
and impact parameters of 0.1, 0.3, 1.0 or 10. With these four impact
parameters most of all possible lightcurve shapes are sampled, as is
illustrated in figure \ref{fig:pacshapes}.  For each simulated
microlensing event, the time sampling and flux error bars were taken
from a random real r$^{\prime}$ lightcurve, based on the 99/00 and 00/01 data. 
In this way, each simulated event
has realistic characteristics, and the total sample of simulated
events has the same observing charactistics as the total set of
observed r$^{\prime}$ lightcurves.  Standard Paczynski microlensing lightcurves
were constructed, using the gaussian error bars taken from the real
data lightcurve.

\begin{figure}
\centering
\includegraphics[width=8cm]{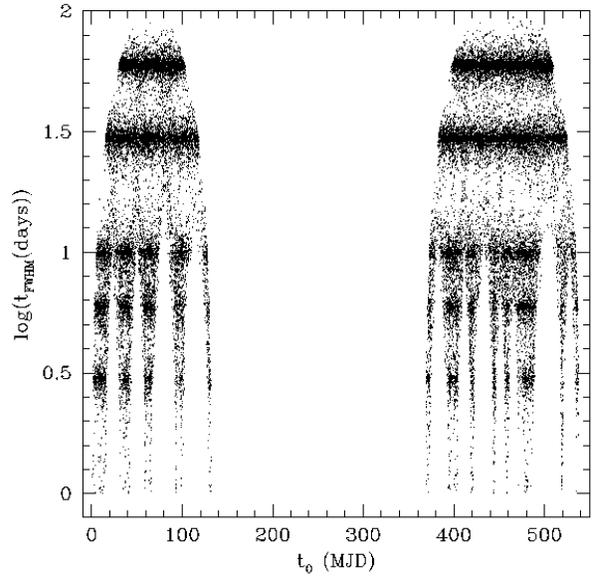}
\caption{For all simulated microlensing lightcurves that pass our
filtering procedure the fitted $t_{FWHM}$ and $t_{max}$ are plotted. From
the clustering of the points it is clear that the fit program manages
to recover the input widths of the simulations quite well. The gaps in
the lightcurves show up as the events get shorter, while the period
during which longer events can be detected is shorter than for the
shorter events.}
\label{fig:mulsimtft0}
\end{figure}

\begin{table}
\begin{center}
\begin{tabular}{l|ccccccc}
\hline
\hline
$t_{FWHM}$ & \multicolumn{7}{l}{Peak flux (ADU/s)} \\
(days) & 2 & 3.5 & 5 & 7 & 10 & 15 & 50 \\
\hline
60 & 47 & 86 & 93 & 95 & 97 & 97 & 98\\
30 & 19 & 71 & 87 & 95 & 98 & 99 & 99\\
10 & 4 & 22 & 38 & 58 & 68 & 74 & 82\\
6 & 1 & 10 & 24 & 35 & 46 & 54 & 61\\
3 & 0 & 1 & 9 & 20 & 27 & 32 & 40\\
1 & 0 & 0 & 0 & 0 & 2 & 4 & 5\\
\hline
\end{tabular}
\begin{small}
\caption{The efficiencies of our r$^{\prime}$ lightcurve filtering procedure, in percent, for the simulated microlensing lightcurves.}
\label{tab:muleff}
\end{small}
\end{center}
\end{table}

In table \ref{tab:muleff} the efficiencies of the selection of
microlensing lightcurves are tabulated for all combinations of
$t_{FWHM}$ and peak flux of the simulated events. For these detectability
calculations simulated microlensing events were used for which the
part of the peak higher than half of the peak flux lay completely
within one observing season, which is one of our selecion
criteria. It is clear that events with a peak flux lower than 3
ADU/s will hardly be selected. Also, events with a
$t_{FWHM}$ lower than 10 days are relatively problematic. This large
decrease in detectability for short timescale events is
primarily caused by gaps in the time sampling of our lightcurves.
This is shown in figure \ref{fig:mulsimtft0}, where the fitted $t_{FWHM}$
and $t_{max}$ of the simulated microlensing lightcurves that pass the
filtering procedure are plotted.

Since long period variable stars like Mira's can mimic microlensing
lightcurves, it is important to determine to what extent these
lightcurves are selected by our microlensing selection
procedures. Although Miras do not have a flat baseline on a
logarithmic plot, on a linear flux scale the peaks are much sharper
and the minima very near zero, thus they are typically below our noise
level except for short periods around the peak. 
Mira variables have periods ranging from 130 to 500 days,
with the most typical period being 270 days (Petit \cite{varstars}).
However, there are a number of Mira's known with even longer periods
(e.g. Jura et al. \cite{Jura93}; Rosino et al. \cite{Rosino97}).
Monte Carlo simulations of 120,000 Mira-type lightcurves
were made with periods ranging from 150 to 550 days. The
lightcurves were assumed to have a symmetric sawtooth shape in
magnitude, with intrinsic variability amplitudes of 2, 4 or 6
magnitudes and 5 different flux variability amplitudes. Time sampling
and flux errors were taken from real data lightcurves in the same way
as for the microlensing simulations and starting phase of the
variability was chosen randomly.

\begin{figure}
\centering
\includegraphics[width=8cm]{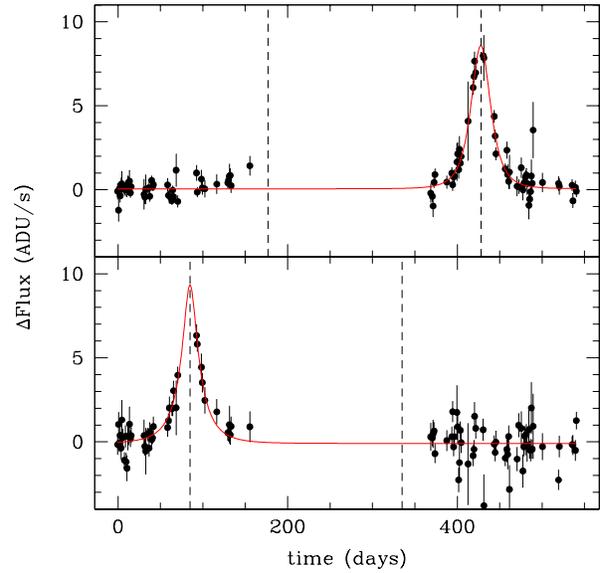}
\caption{Two examples of simulated Mira lightcurves with periods of
250 days that pass our filters. The solid curves show the Paczynski
fit to the lightcurves, and the dashed vertical lines indicate when
the simulated Mira peaks. Both of the lightcurves have peaks in only
one of the observing seasons, and therefore mimic microlensing quite
well. A third observing season solves this problem for variables with
periods around 250 to 300 days, but not very extremely long periods,
i.e. longer than 500 days. Trends in the baseline like at the end of
the first season in the upper lightcurve are often stronger in
i$^{\prime}$ band than in r$^{\prime}$ band.}
\label{fig:miras}
\end{figure}

The filter efficiencies for the selection of simulated Mira
lightcurves are tabulated in table \ref{tab:mireff}. Variable stars
with periods up to 200 days are not likely to be selected and confused
for microlensing events, because at least two peaks will always be
present in the lightcurve. Periods between 200 and 350 days are
clearly a problem for our 2 year dataset. This is caused by the time coverage
of our lightcurves, which for these periods often gives a peak in one
of the observing seasons and a second peak just in between the
seasons.  This is why using the data from the third season is
important; variables with these periods that show only one peak in a
two season lightcurve will show a second peak in the third season. Two
examples of simulated Mira lightcurves with periods of 250 days that
mimic microlensing and pass our filters are shown in figure
\ref{fig:miras}.  Periodic variables with periods around 350 to 450
days are not selected because the fits are never very good and the
baselines not flat. This is because a second peak will always have
wings extending into one of the observing seasons. Longer period
variables, however, can have only one peak in the lightcurve and a
baseline that seems flat. Unfortunately, even with the third observing
season data, these variables can still show one single peak and an
otherwise flat baseline. Other characteristics, like colour and peak
shape, have to be used to distinguish these variables from genuine
microlensing events.  For variable stars of all periods, the highest
amplitude variables are least likely to be mistaken for microlensing
events, because the difference in shape of the peak is detected better
with higher amplitude.

The $t_{FWHM}$ values of the simulated Miras show that variables with
longer periods have broader peaks, as can be expected. However, even
the longest timescale simulated Miras show average $t_{FWHM}$ values
between 30 and 40, while the real variables cluster around a
$t_{FWHM}$ of 50 (Figure \ref{fig:tfcolour}). Most detected real
variables turn out to have more irregular peak shapes and lower peak
heights than our simulated Miras, so that in reality long period
variable stars are less likely to pass through our filtering procedure
than table \ref{tab:mireff} suggests.

\begin{table}
\begin{center}
\begin{tabular}{l|ccccc}
\hline
\hline
Period & \multicolumn{5}{l}{Amplitude (ADU/s)} \\
(days) & 5 & 10 & 15 & 20 & 30 \\
\hline
150 & 5 & 2 & 1 & 0 & 0\\
200 & 5 & 3 & 2 & 2 & 1\\
250 & 34 & 26 & 24 & 20 & 17 \\
300 & 25 & 22 & 19 & 17 & 15\\
350 & 2 & 1 & 1 & 1 & 0\\
400 & 0 & 0 & 0 & 0 & 0\\
450 & 6 & 4 & 3 & 2 & 2\\
500 & 17 & 12 & 10 & 8 & 8\\
550 & 29 & 23 & 21 & 17 & 20\\
\hline
\end{tabular}
\begin{small}
\caption{The efficiencies of our r$^{\prime}$ lightcurve filtering procedure, in percent, for the simulated Mira lightcurves.}
\label{tab:mireff}
\end{small}
\end{center}
\end{table}

\begin{thebibliography}{}
\bibitem[1993]{Alcock93}
	Alcock, C., Akerloff, C.W., Allsman, R.A. et al. 1993, Nature, 365, 621
\bibitem[2000]{Alcock00}
	Alcock, C., Allsman, R.A., Alves, D.R. et al. 2000, ApJ, 542, 281
\bibitem[1993]{Aubourg93}
	Aubourg, E., Bareyre, P., Brehin, S. et al. 1993, Nature, 365, 623
\bibitem[2001]{Auriere01}
	Auri\`ere, M., Baillon, P., Bouquet, A. et al. 2001, ApJ, 553, L137
\bibitem[1993]{Baillon93}
	Baillon, P., Bouquet, A., Giraud-Heraud, Y. \& Kaplan, J. 1993, A\&A, 277, 1
\bibitem[2000]{BaltzSilk00}
	Baltz, E.A. \& Silk, J. 2000, ApJ, 530, 578
\bibitem[2003]{Baltz03}
	Baltz, E.A., Gyuk, G. \& Crotts, A.P.S. 2003, ApJ, 582, 30 (BGC)
\bibitem[1996]{SExtractor}
	Bertin, E. \& Arnouts, S. 1996, A\&AS, 117, 393
\bibitem[1993]{Bottema93}
	Bottema, R. 1993, A\&A, 275, 16
\bibitem[2003]{Calchi03}
	Calchi Novati, S., Jetzer, Ph., Scarpetta, G. et al. 2003, A\&A, 405, 851
\bibitem[1992]{Crotts92}
	Crotts, A.P.S. 1992, ApJ, 399, L43
\bibitem[1996]{CrottsTomaney96}
	Crotts, A.P.S. \& Tomaney, A.B. 1996, ApJ, 473, L87
\bibitem[2001]{MEGA}
	Crotts, A.P.S., Uglesich, R.R., Gyuk, G. \& Tomaney, A.B. 2001, in ASP Conf. Ser. 237: Gravitational Lensing: Recent Progress and Future Go, eds. T.G. Brainerd \& C.S. Kochanek, 243
\bibitem[1999]{Gondolo99}
	Gondolo, P. 1999, ApJ, 510, L29
\bibitem[1996]{Gould96}
	Gould, A. 1996, ApJ, 470, 201
\bibitem[2000]{GyukCrotts00}
	Gyuk, G. \& Crotts, A.P.S. 2000, ApJ, 535, 621
\bibitem[1993]{Jura93}
	Jura, M., Yamamoto, A. \& Kleinmann, S.G. 1993, ApJ. 413, 298
\bibitem[2001]{Kerins01}
	Kerins, E., Carr, B.J., Evans, N.W. et al. 2001, MNRAS, 323, 13
\bibitem[2000]{Lasserre00}
	Lasserre, T., Afonso, C., Albert, J.N. et al. 2000, A\&A, 355, L39
\bibitem[1976]{Lomb}
	Lomb, N.R. 1976, Ap\&SS, 39, 447
\bibitem[2002]{Milsztajn02}
	Milsztajn, A. 2002, Space Science Reviews, 100, 103
\bibitem[1986]{Pacz86}
	Paczynski, B. 1986, ApJ, 304, 1
\bibitem[2002]{Paulin02}
	Paulin-Henriksson, S., Baillon, P., Bouquet, A. et al. 2002, ApJ, 576, L121
\bibitem[2003]{Paulin03}
	Paulin-Henriksson, S., Baillon, P., Bouquet, A. et al. 2003, A\&A, in press, astro-ph/0207025
\bibitem[1987]{varstars}
	Petit, M. 1987, {\it Variable Stars}, John Wiley \& Sons
\bibitem[1992]{numrec}
	Press, W.H., Teukolsky, S.A., Vetterling, W.T. \& Flannery, B.P. 1992, {\it Numerical Recipes in C}, Cambridge University Press
\bibitem[2001]{Riffeser01}
	Riffeser, A., Fliri, J., G\"ossl, C.A. et al. 2001, A\&A, 379, 362
\bibitem[1997]{Rosino97}
	Rosino, L., Ortolani, S., Barbuy, B. \& Bica, E. 1997, MNRAS, 289, 745
\bibitem[1996]{TomaneyCrotts96}
	Tomaney, A.B. \& Crotts, A.P.S. 1996, AJ, 112, 2872
\end{thebibliography}
\end{document}